\documentclass[sigconf,nonacm]{acmart}

\AtBeginDocument{%
  \providecommand\BibTeX{{%
    \normalfont B\kern-0.5em{\scshape i\kern-0.25em b}\kern-0.8em\TeX}}}

\setcopyright{rightsretained}

\usepackage{algorithm}
\usepackage{algorithmic}
\usepackage{breakcites,balance}

\begin{document}

\title[Practice with Graph-based ANN Algorithms on Sparse Data]{Practice with Graph-based ANN Algorithms on Sparse Data: Chi-square Two-tower model, HNSW,  Sign Cauchy Projections}

\author{\vspace{0.2in}Ping Li, Weijie Zhao, Chao Wang, Qi Xia, Alice Wu, Lijun Peng\vspace{0.05in}}
\affiliation{\institution{LinkedIn Ads\vspace{0.05in}\\
\{pinli, weijzhao, cowang, qixia, aiwu, lpeng\}@linkedin.com \vspace{0.2in}}\country{}}

\begin{abstract}

\footnote{This work was initially presented in Microsoft Research in Feb 2023.}Sparse data are common. The traditional ``handcrafted'' features are often sparse. Embedding vectors from trained models can also be very sparse, for example, embeddings trained via the ``ReLu'' activation function. In this paper, we report our exploration of efficient search in sparse data with graph-based ANN algorithms (e.g., HNSW, or SONG which is the GPU version of HNSW), which are popular in industrial practice, e.g., search and ads (advertising).

\vspace{0.1in}

\noindent We experiment with the proprietary ads targeting application, as well as benchmark public datasets. For ads targeting, we train embeddings with the standard ``cosine two-tower'' model and we also develop the ``chi-square two-tower'' model. Both models produce (highly) sparse embeddings when they are integrated with the ``ReLu'' activation function. In EBR (embedding-based retrieval) applications, after we the embeddings are trained, the next crucial task is the approximate near neighbor (ANN) search for serving. While there are many ANN algorithms we can choose from, in this study, we focus on the graph-based ANN algorithm (e.g., HNSW-type).

\vspace{0.1in}
\noindent Sparse embeddings should help improve the efficiency of EBR. One  benefit is the reduced memory cost for the embeddings. The other obvious benefit is the reduced computational time for evaluating  similarities, because, for graph-based ANN algorithms such as HNSW, computing similarities  is often the dominating cost. In addition to the effort on leveraging data sparsity for storage and computation, we also integrate ``sign cauchy random projections'' (SignCRP) to hash vectors to bits, to further reduce the memory cost and speed up the ANN search. In NIPS'13, SignCRP was proposed to hash the chi-square similarity, which is  a well-adopted nonlinear kernel in NLP and computer vision. Therefore, the chi-square two-tower model, SignCRP, and HNSW are now tightly integrated.
\end{abstract}

\keywords{chi-square similarity, two-tower model, sparse data, graph-based approximate near neighbor search, sign cauchy random projections\vspace{0.2in}}

\maketitle


\section{Introduction}

Embedding has become the standard component in deep learning. Embedding models including BERT~\citep{devlin2019bert}, GLOVE~\citep{pennington2014glove}, GPT-3~\citep{brown2020language} etc. have been widely adopted in practice in NLP, knowledge graphs, computer vision, information retrieval, etc.~\citep{karpathy2014deep,yu2018hierarchical,huang2019knowledge,chang2020pre,giorgi2021declutr,yu2022egm,spillo2022knowledge,lanchantin2021general,neelakantan2022text}.
The two-tower model~\citep{huang2013learning} has become the standard neural architecture for generating embeddings which can be subsequently used for retrievals and other downstream applications. The two-tower model is the foundation for the ``embedding-based retrieval'' (EBR), which has been  broadly adopted in the (search and advertising) industry, for example, for efficiently generating quality candidates as input to (e.g.,) the subsequent advertisements (ads) ranking algorithm~\citep{fan2019mobius} or video recommendation algorithm~\citep{covington2016deep}.

\begin{figure}[h]


\mbox{ \hspace{-0.2in}
    \includegraphics[width=1.71in]{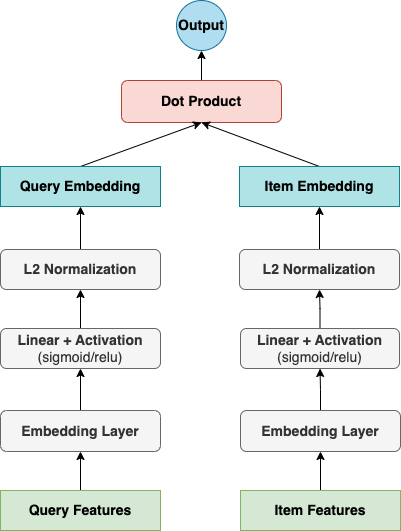}\hspace{0.06in}
    \includegraphics[width=1.71in]{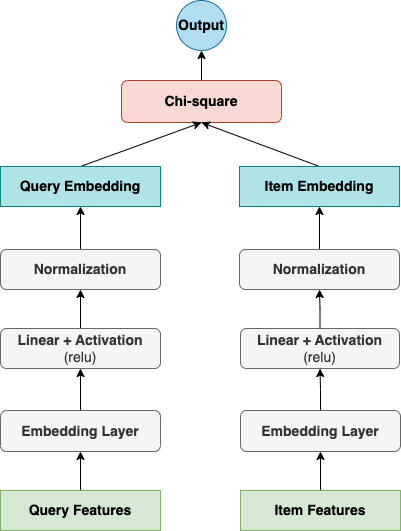}
}


    \caption{Left: the cosine two-tower model. Right: the chi-square two-tower model (which requires to use ReLU).}
    \label{fig:two-tower}
\end{figure}

Figure~\ref{fig:two-tower} (left panel)  provides a simplified illustration of the two-tower model. The top-layer of two-tower model computes the inner product of the two embeddings (which are typically normalized to have the unit $l_2$ norm) of the two towers.  After the model has been trained, the embeddings can be directly used for matching: given a query embedding, searching for the item embeddings with the highest cosine similarity to the query embedding. In recent years, graph-based ANN methods such as HNSW~\citep{malkov2020efficient} or SONG~\citep{zhao2020song}, which is the GPU version of HNSW, are often adopted to speed up the search process.

The cosine similarity between two $d$-dimensional embeddings (vectors) $u,v\in\mathbb{R}^d$ is  defined as
\begin{align}
\rho = \sum_{i=1}^d u_iv_i, \hspace{0.2in}
\sum_{i=1}^d u_i^2 = \sum_{i=1}^d v_i^2 = 1,
\end{align}
if we assume the vectors are  pre-normalized to have the unit $l_2$ norm.  In this study, we propose the ``chi-square two-tower'' model which only slightly modifies the common ``cosine two-tower'' model, by replacing the inner product with the  following chi-square similarity:
\begin{align}\label{eqn:chi-square}
\rho_{\chi^2} = \sum_{i=1}^d \frac{2u_iv_i}{u_i+v_i}, \hspace{0.2in} u_i\geq 0,\hspace{0.1in} v_i\geq 0, \hspace{0.1in}
\sum_{i=1}^d u_i = \sum_{i=1}^d v_i = 1
\end{align}
As shown in the right panel of Figure~\ref{fig:two-tower}, in combination with the ``ReLU'' activation function~\citep{fukushima1975cognitron,nair2010rectified,glorot2011deep}, the embedding vectors are naturally non-negative. The sum-to-one constraint is easy to be enforced in the neural network training process. In fact, the sum-to-one constraint is the same as the $l_1$ (lasso) constraint in non-negative data~\citep{tibshirani1996regression}. When we use the ReLU activation, we observe that both the cosine two-tower model and the chi-square two-tower model produce  sparse embeddings, as expected~\citep{fukushima1975cognitron,nair2010rectified,glorot2011deep}. We are aware of other existing works which aim to produce sparse embeddings~\citep{trifonov2018learning,liang2021anchor}.

There are notable advantages with sparse embeddings. The storage cost can be substantially reduced if the embeddings are highly sparse. For example, if the sparsity (i.e., the number of nonzero entries divided by the vector length) is $<10\sim 20\%$, we can expect a considerable saving in storage using a sparse format. Also, we expect the cost for similarity computations can be substantially reduced if the embeddings are highly sparse. We will illustrate these advantages via extensive experiments on HNSW for fast approximate near neighbor  search. After the graph is built, the major cost of HNSW is spent on computing similarities on the fly while walking on  the graph to search for nodes with highest similarities.

The chi-square similarity~\eqref{eqn:chi-square}, as a nonlinear kernel,  is popular in NLP and computer vision~\citep{chapelle1999support, hein2005hilbertian, jiang2007towards, wang2009building, alexe2010what, vempati2010generalized, vedaldi2012efficient}. Typically those applications used the ``chi-square distance'' $d_{\chi^2}$ instead of $\rho_{\chi^2}$:
\begin{align}\notag
d_{\chi^2} =& \sum_{i=1}^d \frac{(u_i-v_i)^2}{u_i+v_i} = \sum_{i=1}^d \frac{(u_i+v_i)^2-4u_iv_i}{u_i+v_i} \\\notag
=& \sum_{i=1}^du_i+\sum_{i=1}^dv_i - 2\sum_{i=1}^d\frac{2u_iv_i}{u_i+v+i}=2-2\rho_{\chi^2}
\end{align}
since $\sum_{i=1}^d u_i = \sum_{i=1}^dv_i =1$.
In this study, it is not our major focus to demonstrate the advantage of the chi-square over the cosine. In the experiments, we will  show that the chi-square two-tower model produces even more sparse embeddings than the cosine two-tower model, and the chi-square similarity  achieves similar (in some cases even better) accuracy (recall, classification error, etc).  We apply the chi-square two-tower model for a proprietary ads targeting task, and we also conduct HNSW experiments on   public  datasets.

To conclude this introduction section, we shall point out the well-known limitation of the standard two-tower model. This model does not sufficiently consider interactions between queries and items. The rising field of ``neural ranking'' is a  promising direction~\citep{zhu2018learning,zhu2019joint,tan2020fast,zhuo2020learning,gao2020deep,tan2021fast,yu2022egm,zhao2022guitar}, which however does not have the convenience of the two-tower model, because neural ranking models cannot directly use the popular ANN algorithms. In the industry practice, the two-tower model is still very popular especially as a retrieval model,  and practitioners should be aware of the limitations.

\vspace{0.1in}

Next, we review HNSW and the graph-based ANN search.

\section{Graph-based ANN Search}\label{sec:hnsw}

\vspace{0.1in}

\textbf{Graph index.} Graph-based ANN search constructs a proximity graph as index, where a vertex in the graph corresponds to an item embedding and an edge connects two ``neighboring vertices''. The neighborhood relationship is defined on various constraints to make the graph index applicable for the ANN problem.
For instance, graph constraints like Delaunay Graphs~\citep{aurenhammer1991voronoi} ensure that there exists a path with monotonic increasing similarity to the query embedding from any starting vertex. NSW~\citep{malkov2014approximate}, NSG~\citep{fu2019fast},  HNSW~\citep{malkov2020efficient} and SONG~\citep{zhao2020song} (which is the GPU version of HNSW) approximate the Delaunay Graph to reduce the proximity graph construction complexity to subquadratic time. The graph construction is realized by inserting vertices iteratively, i.e., finding the nearest neighbors using the graph searching methods on the current graph and connecting the new vertex with them. Recently, there have been a wide range of research activities based on HNSW, for example, efficiently maintaining/updating the graphs~\citep{xu2022proximity}, integrating constraints (such as geo-filtering or other filtering mechanisms) with ANN search~\citep{zhao2022constrained}, using HNSW for the maximum inner product search (MIPS)~\citep{morozov2018non,zhou2019mobius,tan2021norm}, etc.

\vspace{0.1in}

\begin{algorithm}[h]
\begin{algorithmic}[1]
\STATE Initialize a priority queue $q$; $\textit{res} \leftarrow \emptyset$
\STATE $q.\textit{insert}(\textit{start\_vertex},\textit{similarity}(\textit{start\_vertex},\textit{query}))$
\WHILE{\textbf{true}}
    \STATE $\textit{curr\_idx, curr\_similarity} \leftarrow q.\textit{pop\_highest\_similarity}()$
    \IF{$\textit{curr\_similarity}$ cannot improve the $L^{\textit{th}}$ similarity in \textit{res}}
        \STATE \textbf{break}
    \ENDIF
    \STATE $\textit{res} \leftarrow \textit{res} \cup \{(curr\_idx, curr\_similarity)\}$
    \FOR{\textit{each } neighbor $x$ of $\textit{curr}$}
        \IF{x is not visited}
            \STATE $q.\textit{insert}(x,\textit{similarity}(x,\textit{query}))$
        \ENDIF
    \ENDFOR
\ENDWHILE
\RETURN \textit{res}
\end{algorithmic}
\caption{Graph Searching Algorithm}\label{alg:search}
\end{algorithm}

\noindent\textbf{Graph searching.}
Once a graph is (partially) constructed, ANN search can be performed by traversing the graph. At each step, the algorithm chooses the neighbor that is closest to the query point and continues the search in that direction. This process is repeated until a stopping criterion is met, i.e., no neighbors can improve the current found $L$ nearest neighbors, where $L$ is a parameter that controls the trade-off between searching time. The larger $L$ generates more accurate nearest neighbors but consumes more time. The graph searching is similar to an A* heuristic search~\citep{hart1968formal}, where the priority is the similarity to the query embedding. The details of the searching is illustrated in Algorithm~\ref{alg:search}. A priority queue $q$ is initialized from the similarity of the starting vertex (e.g., vertex $1$) and the query. Then, in each iteration, we extract the vertex $\textit{curr\_idx}$ from the priority queue with the highest similarity, check the stopping criterion, and explore its unvisited neighbors.

\begin{figure}[h!]

\centering

{
    \includegraphics[width=3.3in]{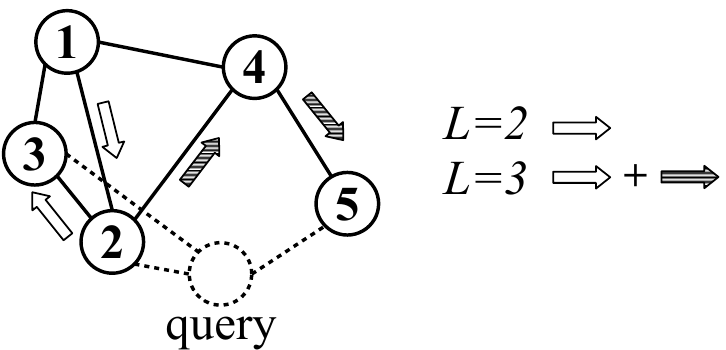}
}

    \caption{Illustration of the graph searching algorithm. When $L=2$, the searching follows the white arrows and stops by returning vertices $2$ and $3$ and results; when $L=3$, the searching continues (following with the hatched arrows) and eventually returns $2$, $5$, and $3$. If our goal is to find top-2 neighbors of the query vector, it might not be sufficient to use $L=2$ (it misses the true neighbor $5$). A greater $L$ helps to return more accurate results.}
    \label{fig:graphsearch}
\end{figure}

Figure~\ref{fig:graphsearch} depicts an example. Consider a query vector at the space of the dotted circle. We start the search from vertex $1$. With $L=2$, the searching stops after exploring vertex $3$ because vertex $4$ cannot improve the currently found top-$2$ results ($2$ and $3$). However, the true top-2 result of the query should be $2$ and $5$. Using a greater $L$, e.g., $L=3$, will let the search continue and successfully locate $5$. Therefore, the $L$ should be chosen properly and should be tuned with the data.

\section{EBR for Ads Targeting}

We illustrate the effectiveness of the chi-square two-tower model using an example from ads targeting, for which we learn member and campaign embeddings from a two-tower model trained on ads engagement tasks, and then use it to retrieve matching campaigns for members based on relevance (similarity) scores computed from those embeddings. For the training, we use various features - member demographics features (e.g., geo location, degree, title), user behavioral features, contextual features and campaign features. All input features first go through a shared embedding layer, where numerical, categorical, textual and id features are mapped to a dense embedding vector. 
We then explore the effectiveness of different similarity measures in the two-tower model training in terms of top-K ranking metrics and retrieval efficiency. The model was trained with 4.5 billion records from data collected on a social network.

\begin{figure}[h]

\mbox{\hspace{-0.25in}
    \includegraphics[width=3.6in]{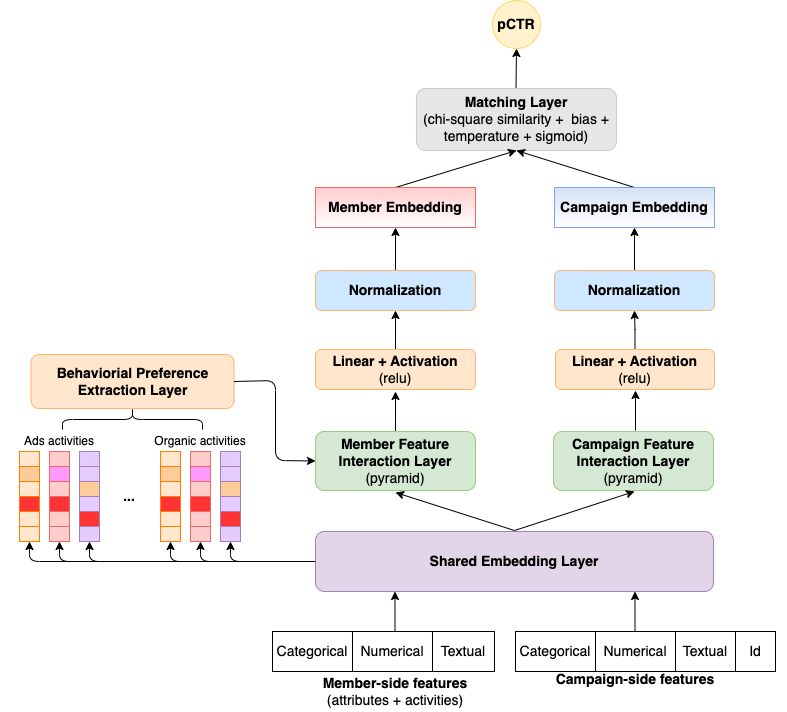}
}

\vspace{-0.05in}

    \caption{The neural architecture for using the chi-square two-tower model for ads targeting.}
    \label{fig:ads}
\end{figure}

Figure~\ref{fig:ads} illustrates the neural architecture for training the embeddings. It is a simple model and  we believe the simplicity allows us to clearly demonstrate the effectiveness of using the chi-square similarity to replace the standard cosine similarity inside the model. Basically, the normalization layer in Figure~\ref{fig:ads} enforces the sum-to-one constraint. The (top) matching layer applies the usual cross-entropy loss with the standard bias and temperature terms. For the comparison, we  use essentially the same architecture to train the cosine two-tower model by replacing the chi-square similarity with cosine and the sum-to-one normalization with the $l_2$ normalization.

\begin{figure}[b!]

\vspace{-0.1in}

\centering
\mbox{

    \includegraphics[width=2.6in]{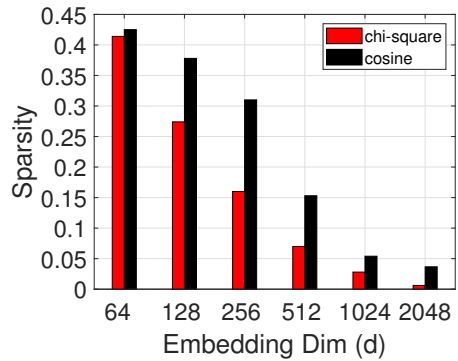},
}

\vspace{-0.1in}

    \caption{Average sparsity values of the embeddings. In this paper, ``sparsity'' is defined as the fraction of non-zero entries, i.e., the number of non-zero entries divided by the embedding dimension.}
    \label{fig:ads_sparsity} 
\end{figure}

Figure~\ref{fig:ads_sparsity} visualizes  the (average) sparsity values for embedding sizes varying from $k=128$ to $k=2048$. As the cosine two-tower model also uses the ReLU activation function, we can see that the model is also quite sparse especially for large embedding size $d$. Nevertheless, the embeddings from the chi-square two-tower model are substantially more sparse. For example, at $d=1024$, the sparsity for the cosine two-tower model is about $5.4\%$, and for  the chi-square two-tower model is about $2.8\%$. In fact, when $d=2048$, the sparsity values are $3.6\%$ and $0.6\%$ respectively for the cosine and the chi-square model. It would be interesting to understand the limit of sparsity by keeping increasing $d$, but we have not conducted such an investigation for $d>2048$. The AUC evaluations are pretty close for both models. For smaller $d$, the AUC scores increase with increasing $d$, and we do not observe improvement after $d\geq 512$.

\begin{figure}[h]


\mbox{\hspace{-0.2in}
    \includegraphics[width=1.85in]{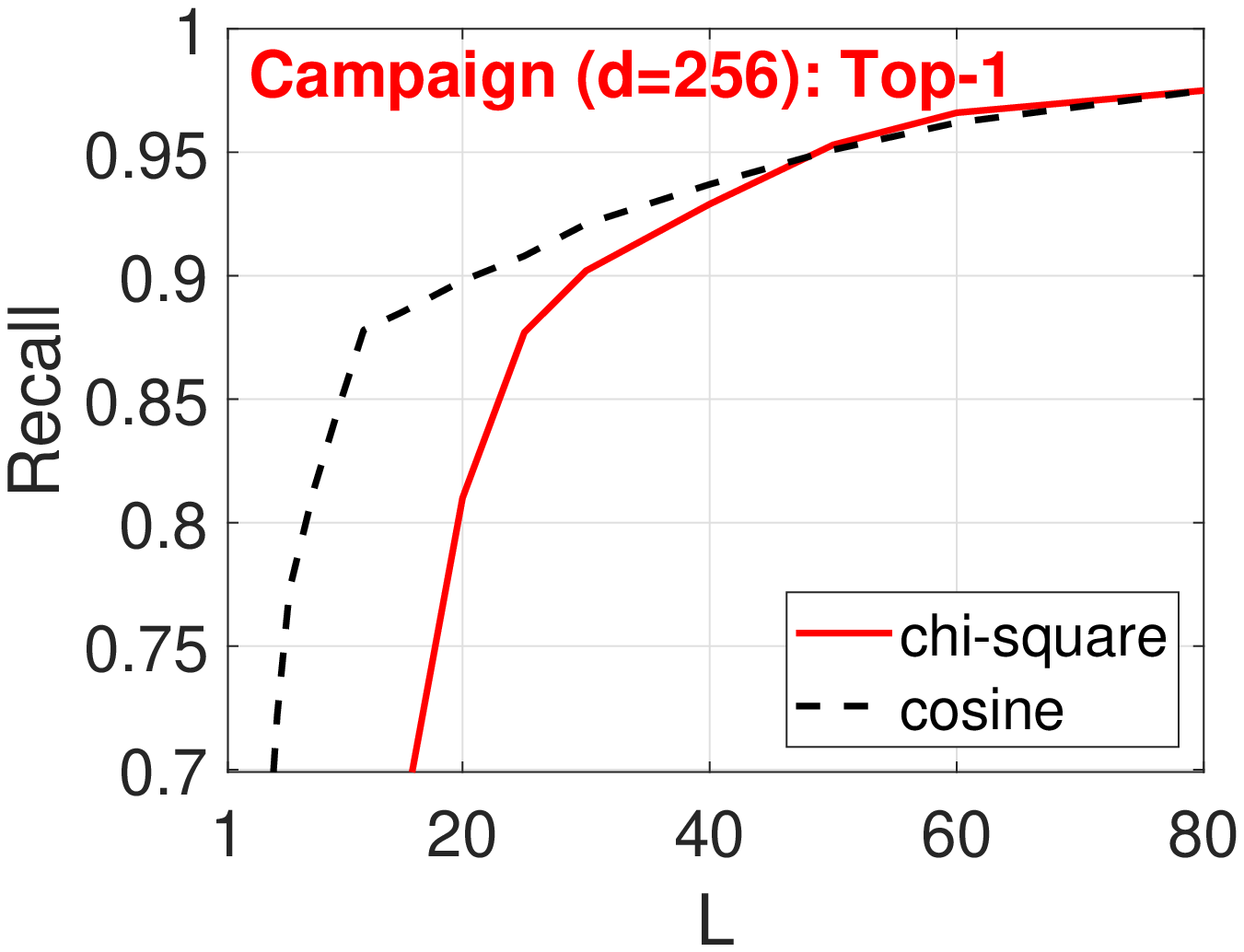}\hspace{-0.1in}
    \includegraphics[width=1.85in]{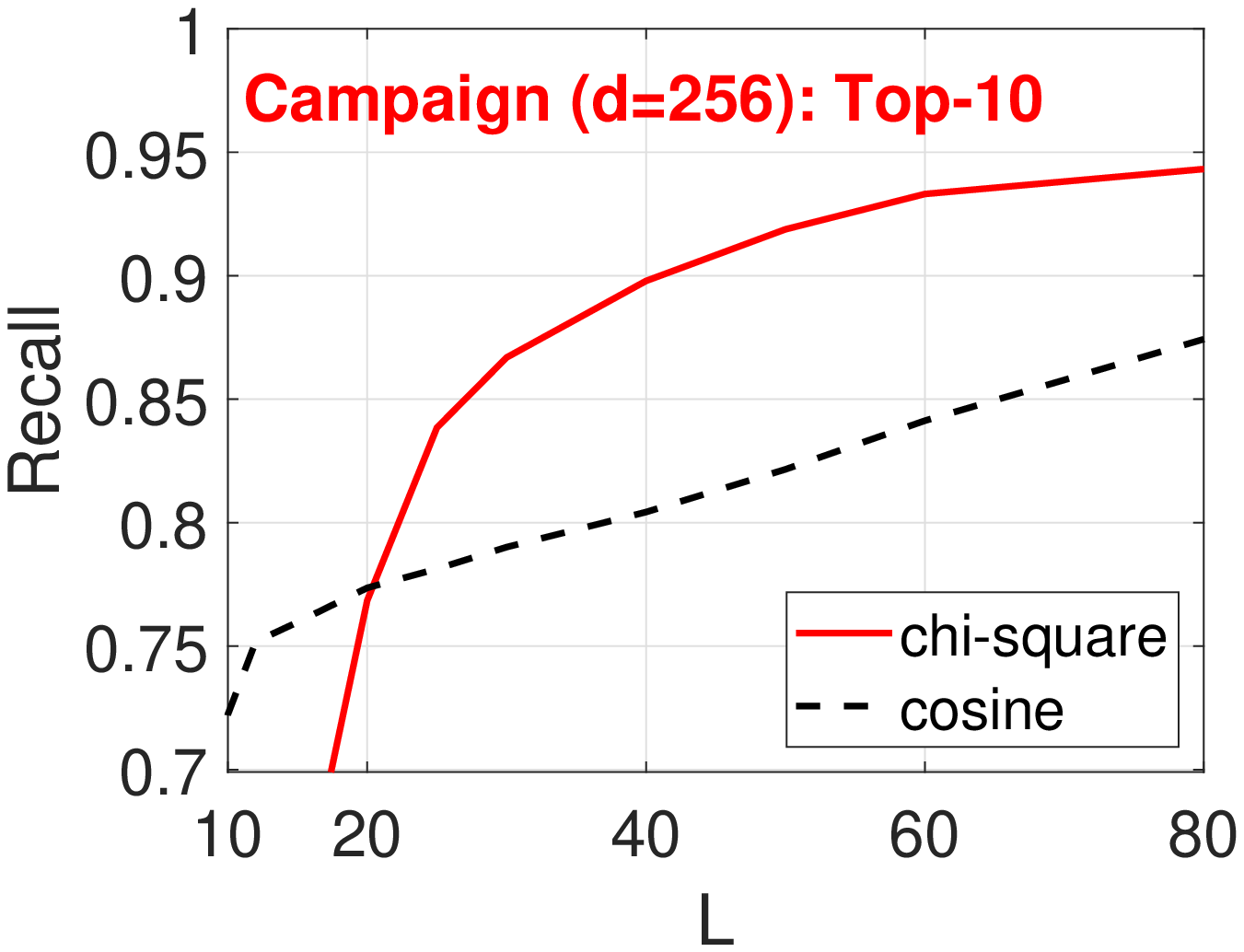}
}

\mbox{\hspace{-0.2in}
    \includegraphics[width=1.85in]{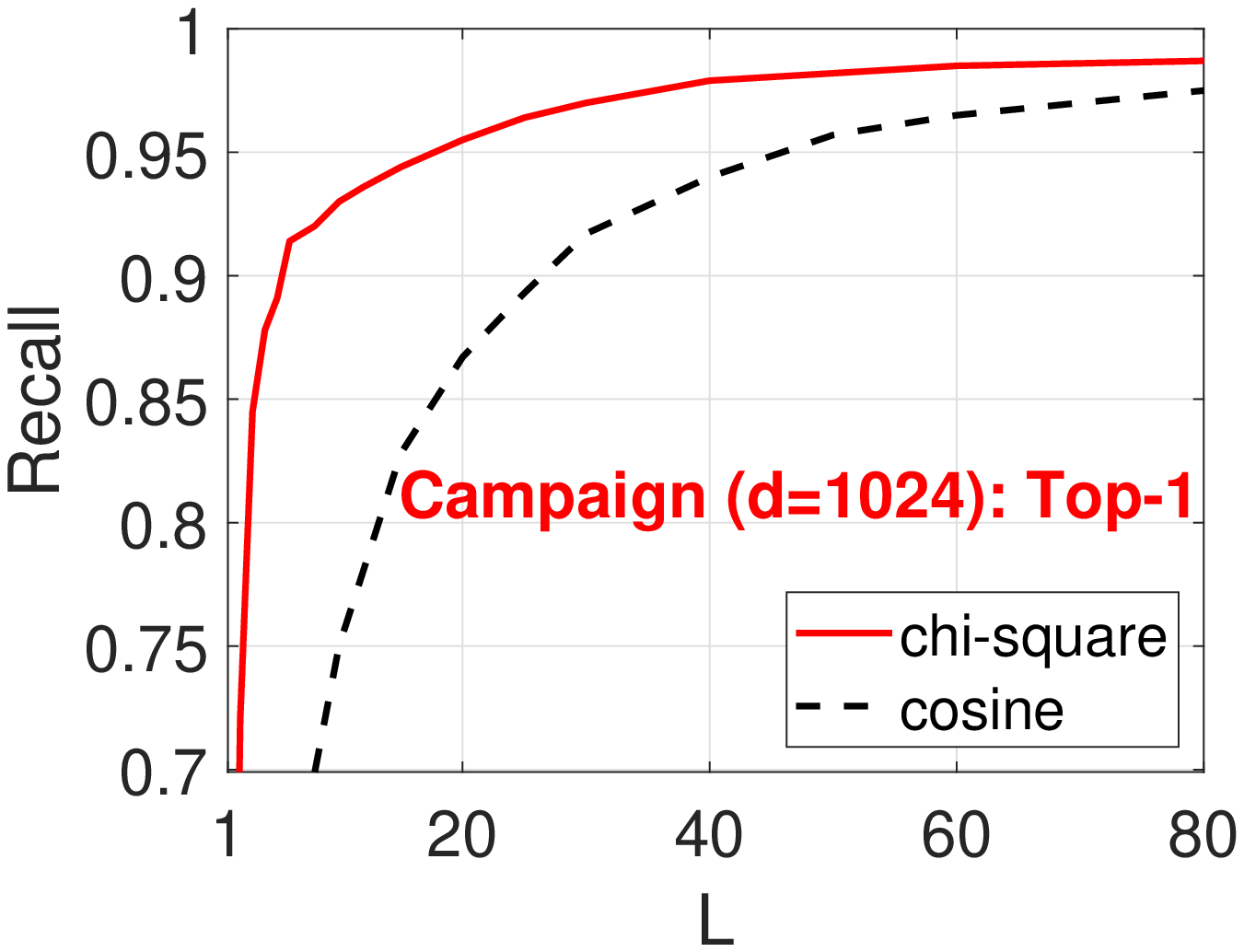}\hspace{-0.1in}
    \includegraphics[width=1.85in]{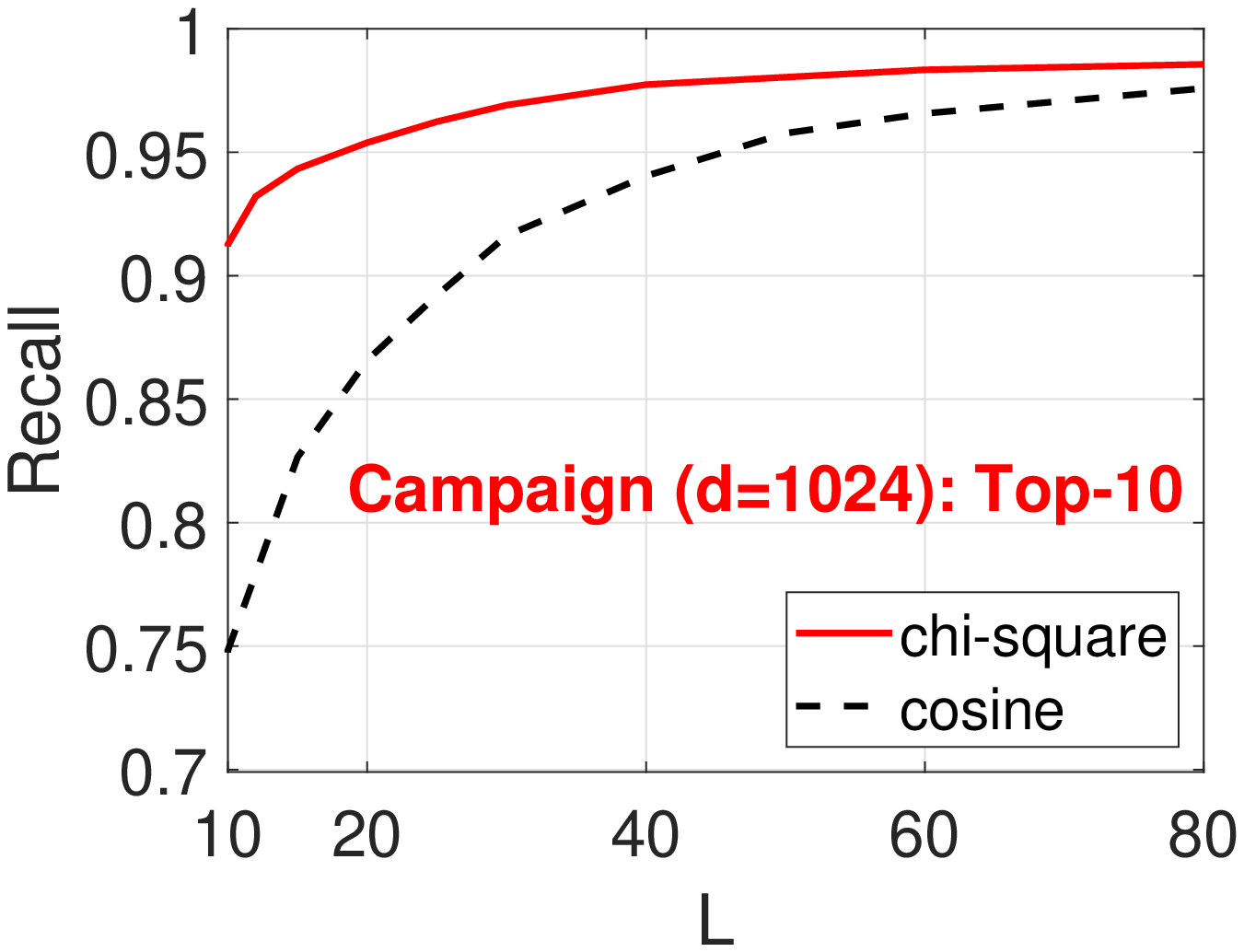}
}

\vspace{-0.05in}

    \caption{Recalls at different $L$ values. Recall in Section~\ref{sec:hnsw}, the searching stops when the current searching vertices cannot improve current found $L$ nearest neighbors.}
    \label{fig:ads_recall}
\end{figure}

After the two-tower models have been trained, we compute the campaign and member embeddings. Then we use member embeddings to search for the most similar campaign embeddings, using the popular graph-based (HNSW) ANN algorithm.  Figure~\ref{fig:ads_recall} reports the top-1 and top-10 recalls for $d=256$ and $d=1024$.  Recall that in HNSW the critical parameter is $L$, which controls the trade-off between the search quality and the search time. We observe from the experiments that for both models, the number of retrieved data points at the same $L$ is very close (also see Figure~\ref{fig:num_retrieved}). We thus use $L$ as a convenient measure for
 comparisons (at the same $L$).

Figure~\ref{fig:ads_recall} confirms the high-quality of HNSW in that it does not need to search many points (i.e., small $L$ values) in order to achieve a reasonable recall say 0.9. Another interesting observation is that using larger embeddings can be actually faster for this application. For example, with $d=1024$, we just need $L=5$ to achieve a top-1 recall of 0.9, while we need $L=30$ to achieve 0.9 if $d=256$.

We should mention that,  with our chi-square model, the number of nonzeros at $d=256$ is actually more than the number of nonzeros at $d=1024$ (i.e., 41 versus 29). This set of experiments suggests that it might be better to use larger and sparser embeddings instead of shorter embeddings, as far as the search quality/efficiency is concerned. We will also report similar results on public data.
\vspace{0.2in}

\begin{figure}[t]


    \centering

\mbox{
    \includegraphics[width=2.6in]{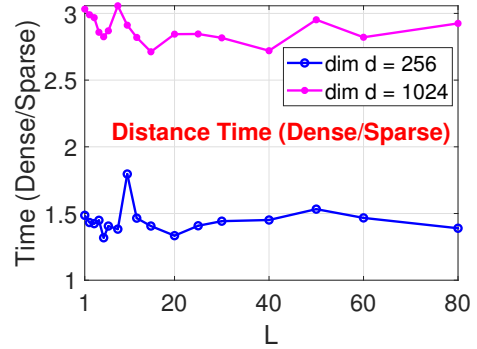}
}

\vspace{-0.1in}

    \caption{Ratios of distance computing times for the chi-square two-tower model: time for using dense format over the time for using sparse format.}
    \label{fig:ads_time}
\end{figure}

Figure~\ref{fig:ads_time} reports the saving in time by using sparse data format in the implementation of HNSW. Typically, in EBR applications, embedding vectors are stored with dense representations. With our chi-square two-tower model, as the sparsity is merely $2.8\%$ at $d=1024$, we obtain a huge saving in memory using sparse format. 
Figure~\ref{fig:ads_time} shows that we achieve a 3-fold reduction in time  for computing similarities by using the sparse format when $d=1024$. 

\vspace{.15in}

\section{Experiments on Public Data}

To further understand the difference between chi-square  and  cosine similarities in the context of approximate near neighbor search, we conduct HNSW experiments on four  public datasets. The dataset specifications are shown in Table~\ref{tab:data}.

\begin{table}[h]


\centering
\caption{Summary statistics of four public datasets.}\label{tab:words}
    \begin{tabular}{crrrrr}
    \hline\hline
        dataset  & \# queries & \# vectors&  \# dim & \# nonzero &Sparsity\\\hline
        Webspam &5,000 &345,000 &16.6M &3,720 & $0.02\%$\\
        News20 &3,993 &15,935 &62,061 &79.9 & $0.13\%$\\
        RCV1 &15,564&518,571 &47,236 &64.6 & $0.14\%$\\
        MNIST &10,000 &60,000 &784 &149.9 & $19.12\%$\\
        \hline\hline
    \end{tabular}\label{tab:data}
\end{table}

\begin{figure}[b!]

\vspace{-0.2in}

\mbox{\hspace{-0.2in}
    \includegraphics[width=1.85in]{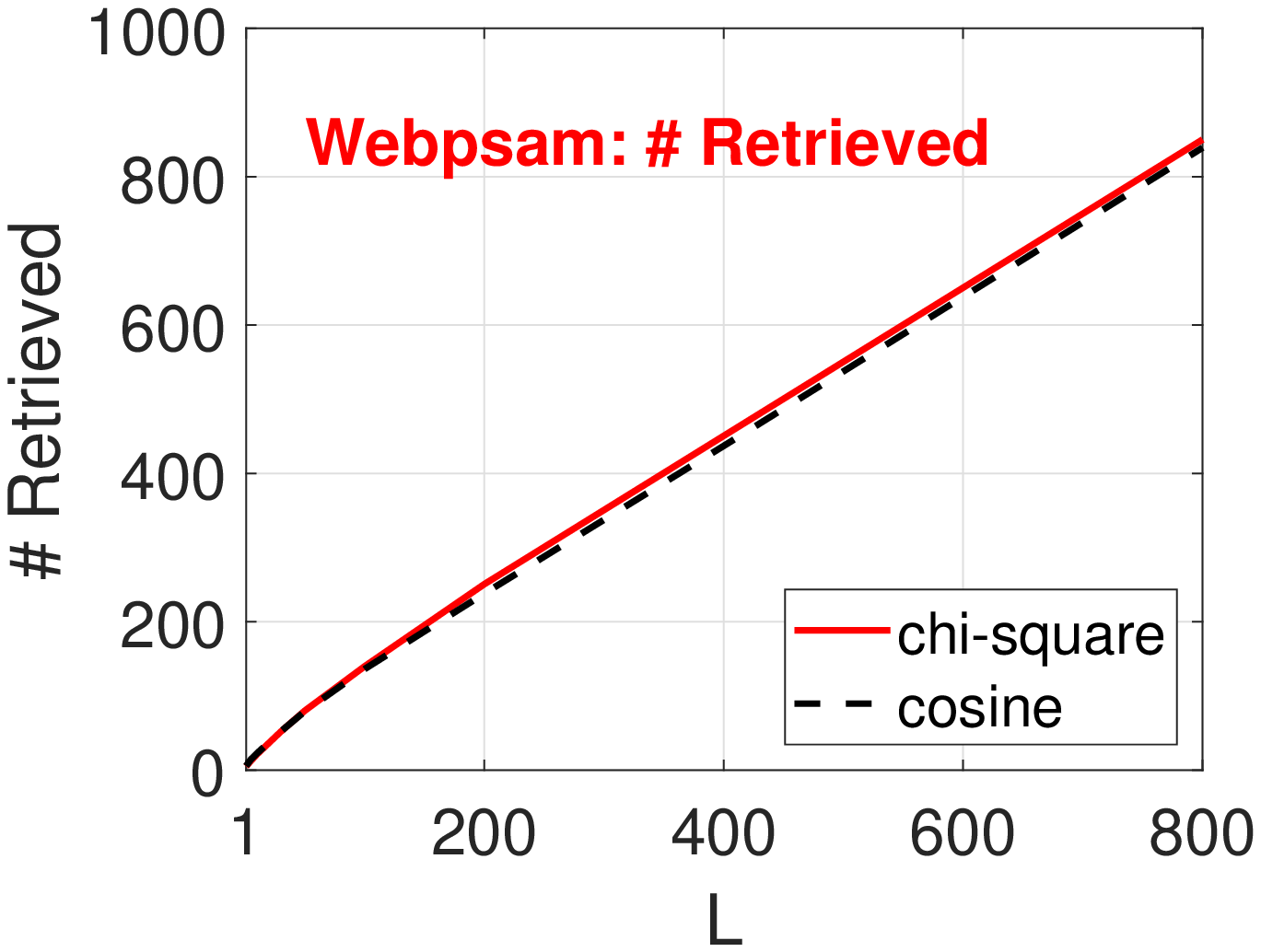}\hspace{-0.1in}
    \includegraphics[width=1.85in]{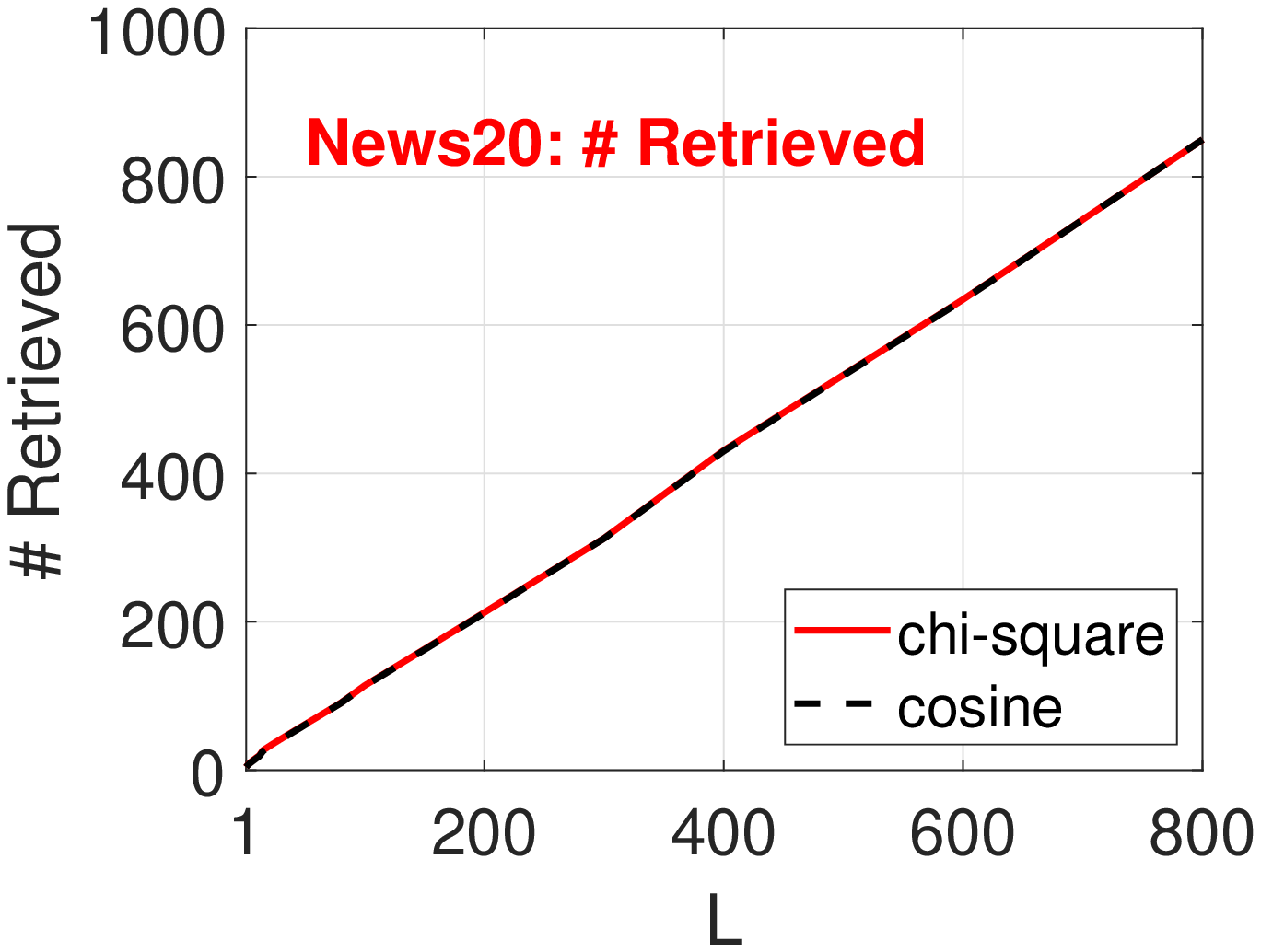}
}

\mbox{\hspace{-0.2in}
    \includegraphics[width=1.85in]{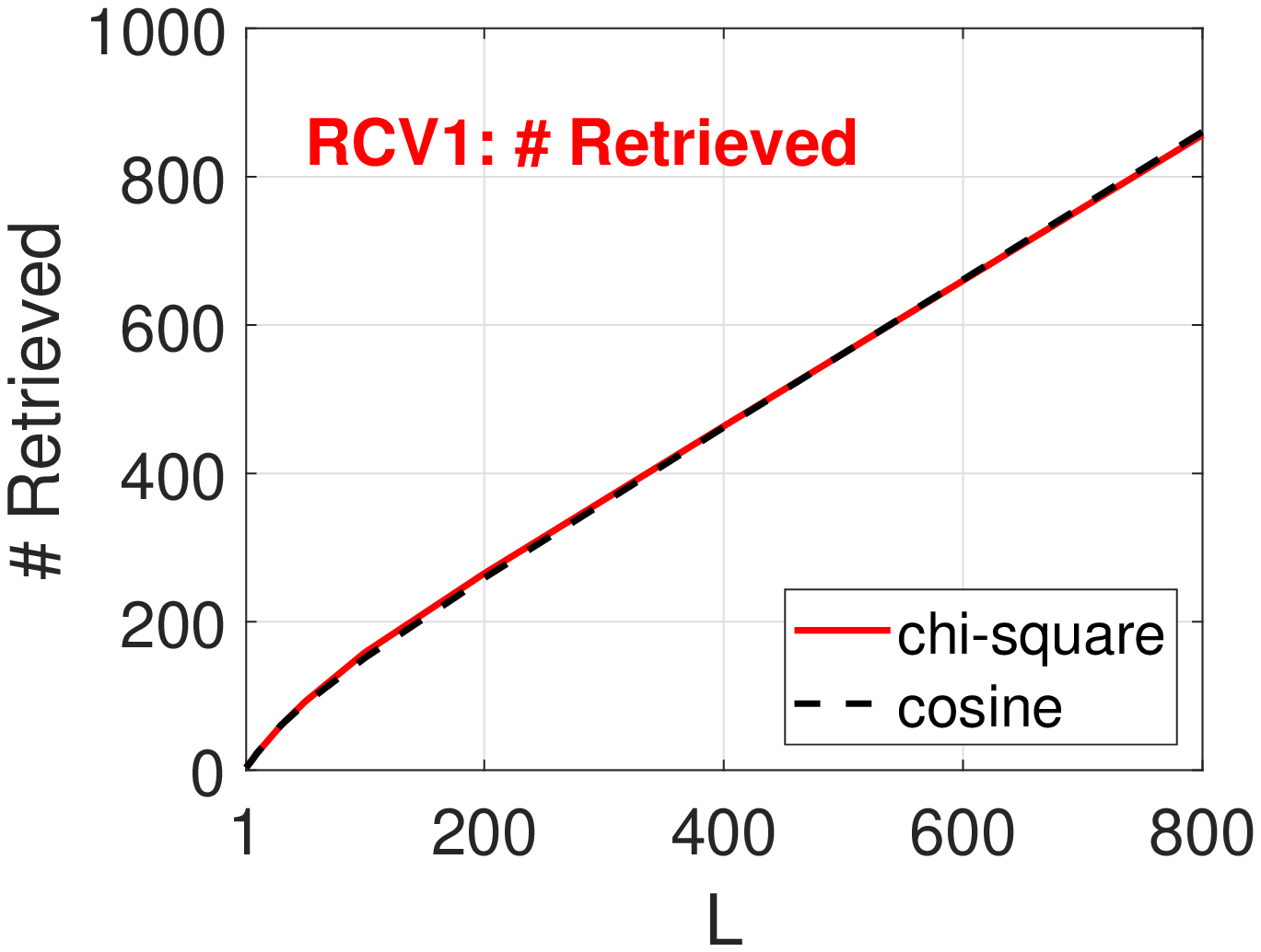}\hspace{-0.1in}
    \includegraphics[width=1.85in]{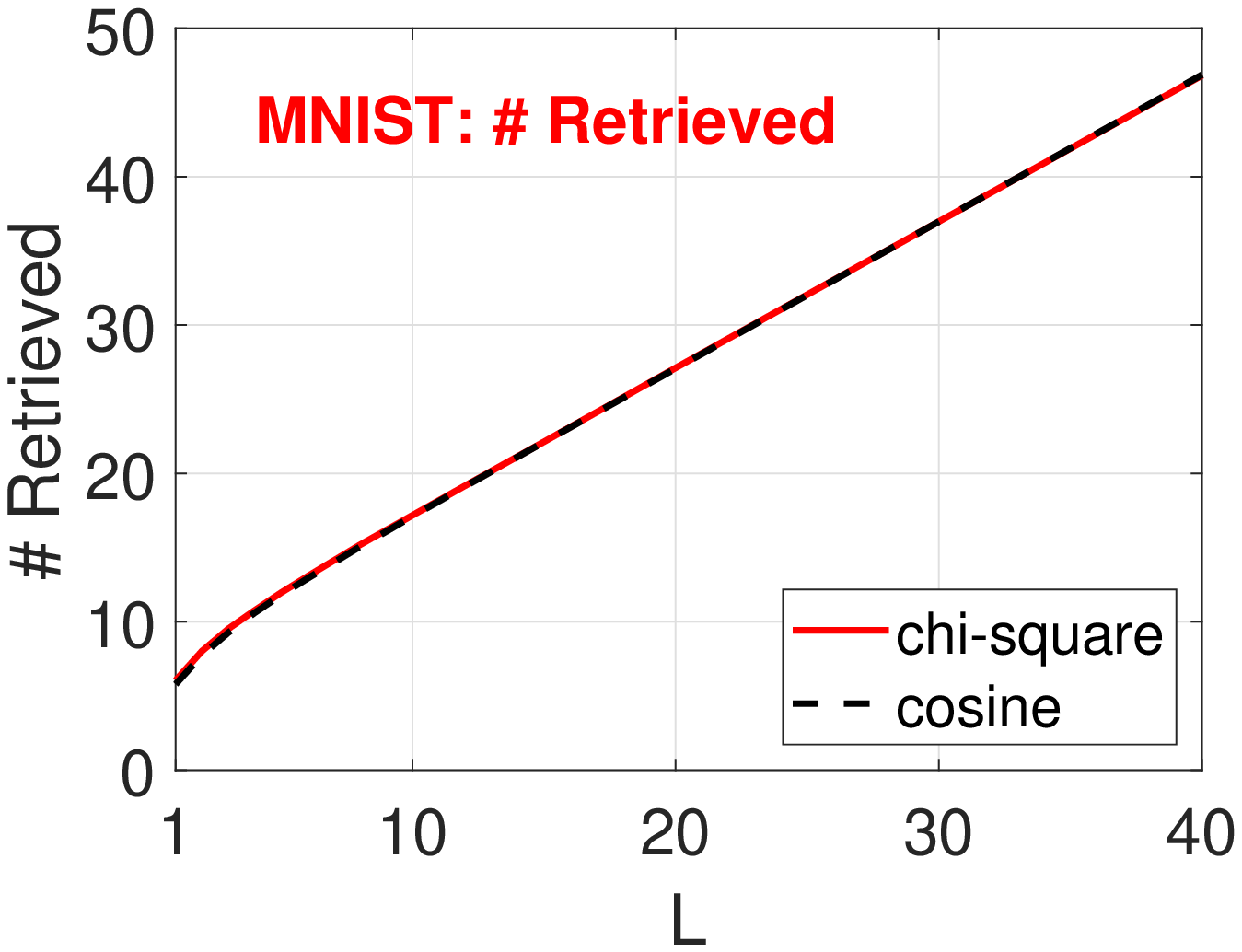}
}

\vspace{-0.05in}

    \caption{The number of retrieved data vectors with respect to the HNSW parameter $L$.}
    \label{fig:num_retrieved}
\end{figure}

For all four datasets, we conduct ANN experiments separately for two different ground-truths: the cosine similarity and the chi-square similarity.
 Figure~\ref{fig:num_retrieved} reports the number of retrieved vectors with respect to the parameter $L$ in HNSW. We can see that the numbers are essentially proportional to $L$ and do not differ much between chi-square and cosine.  Figure~\ref{fig:num_retrieved} confirms that it is appropriate to use $L$ for measuring/comparing the performance of HNSW experiments.

\begin{figure}[h!]


\mbox{\hspace{-0.2in}
    \includegraphics[width=1.85in]{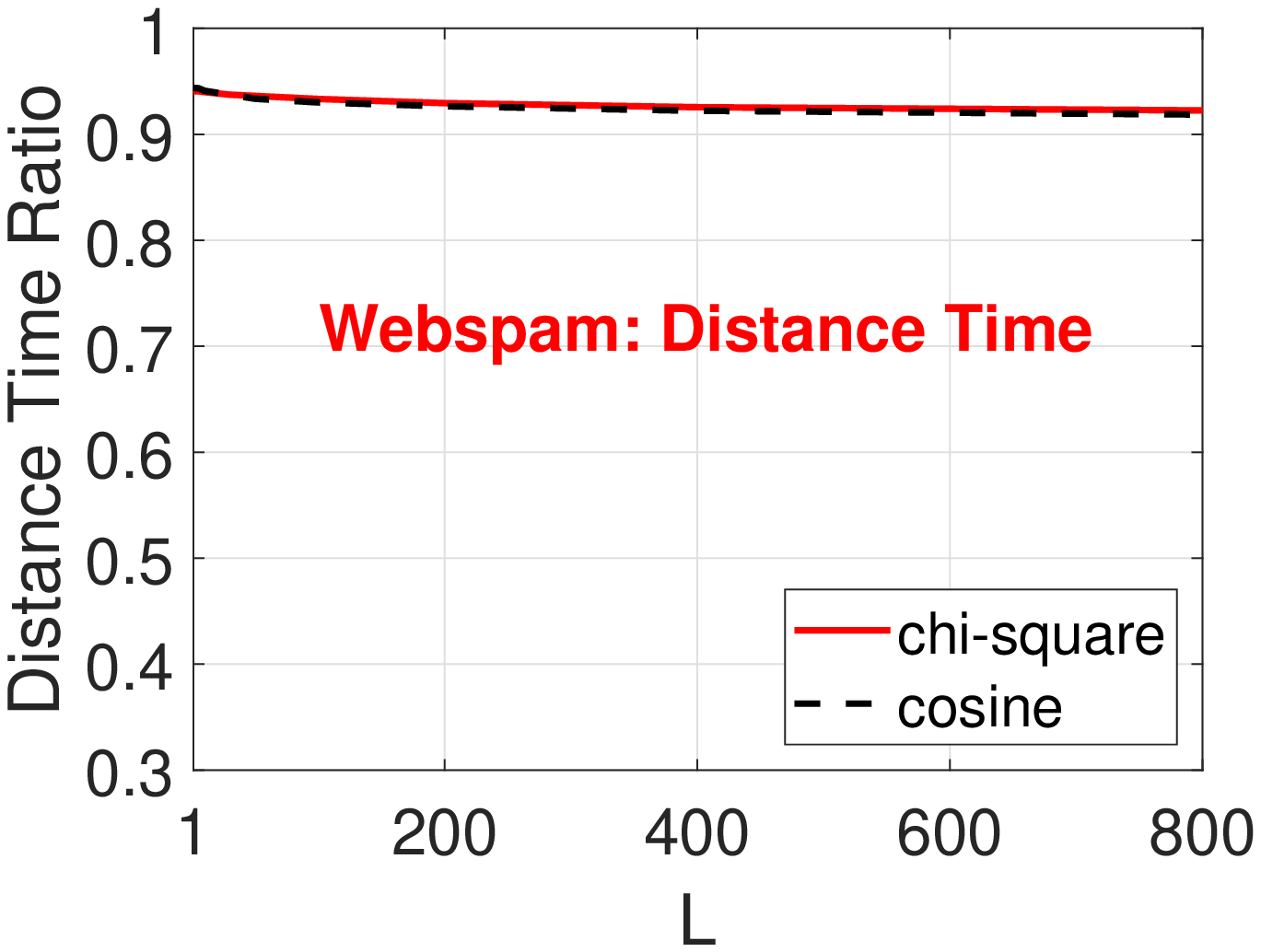}\hspace{-0.1in}
    \includegraphics[width=1.85in]{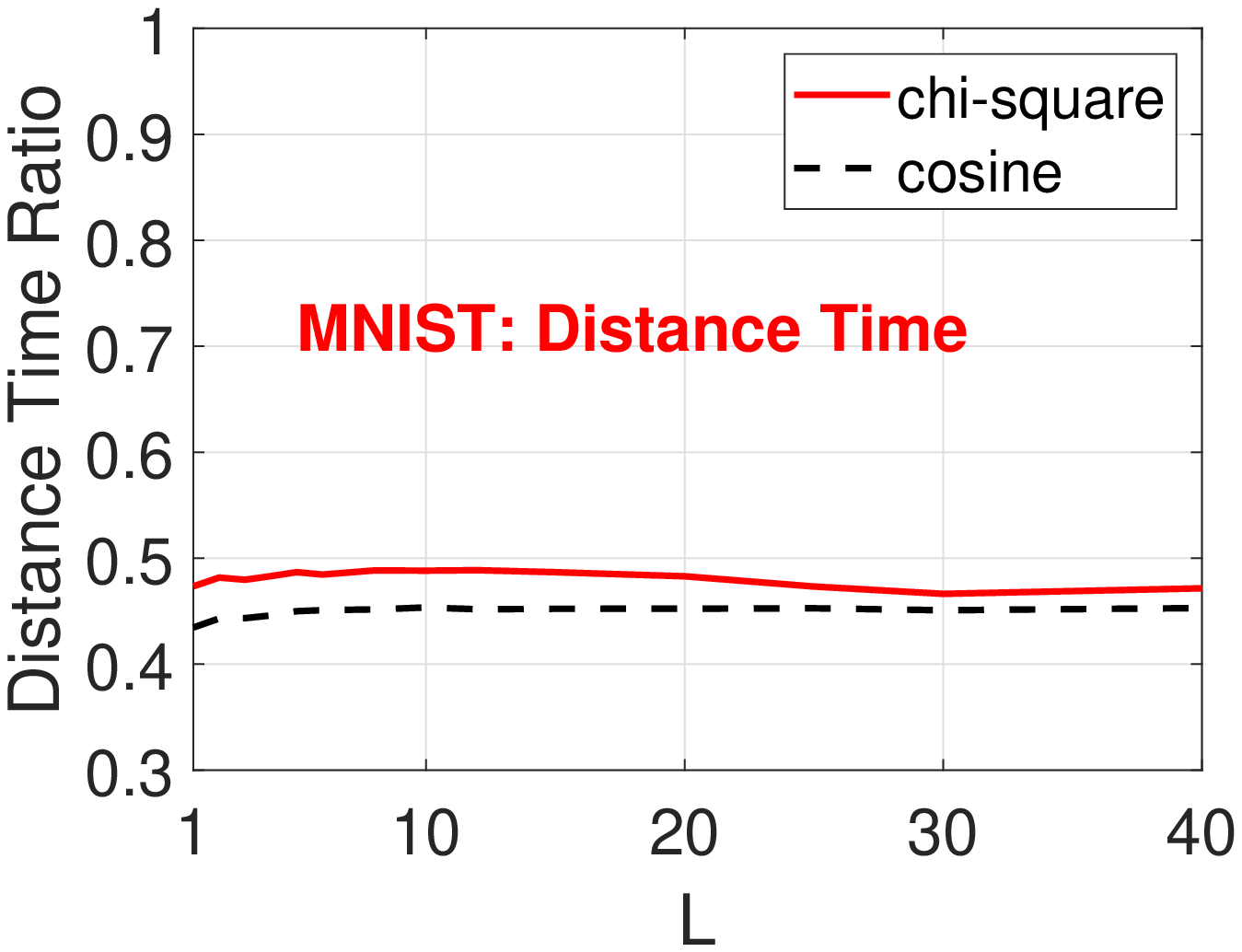}
}

\mbox{\hspace{-0.2in}
\includegraphics[width=1.85in]{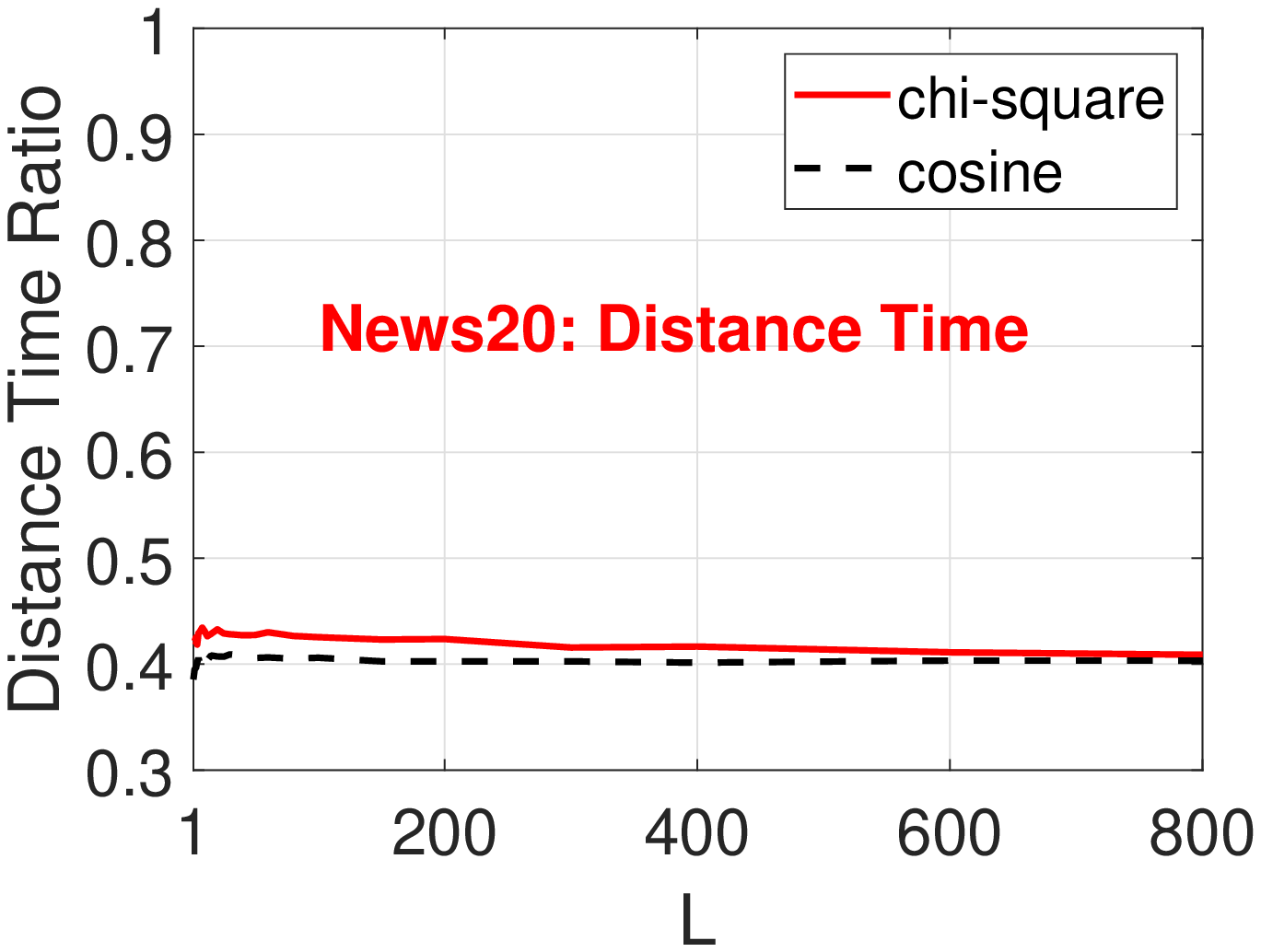}\hspace{-0.1in}
    \includegraphics[width=1.85in]{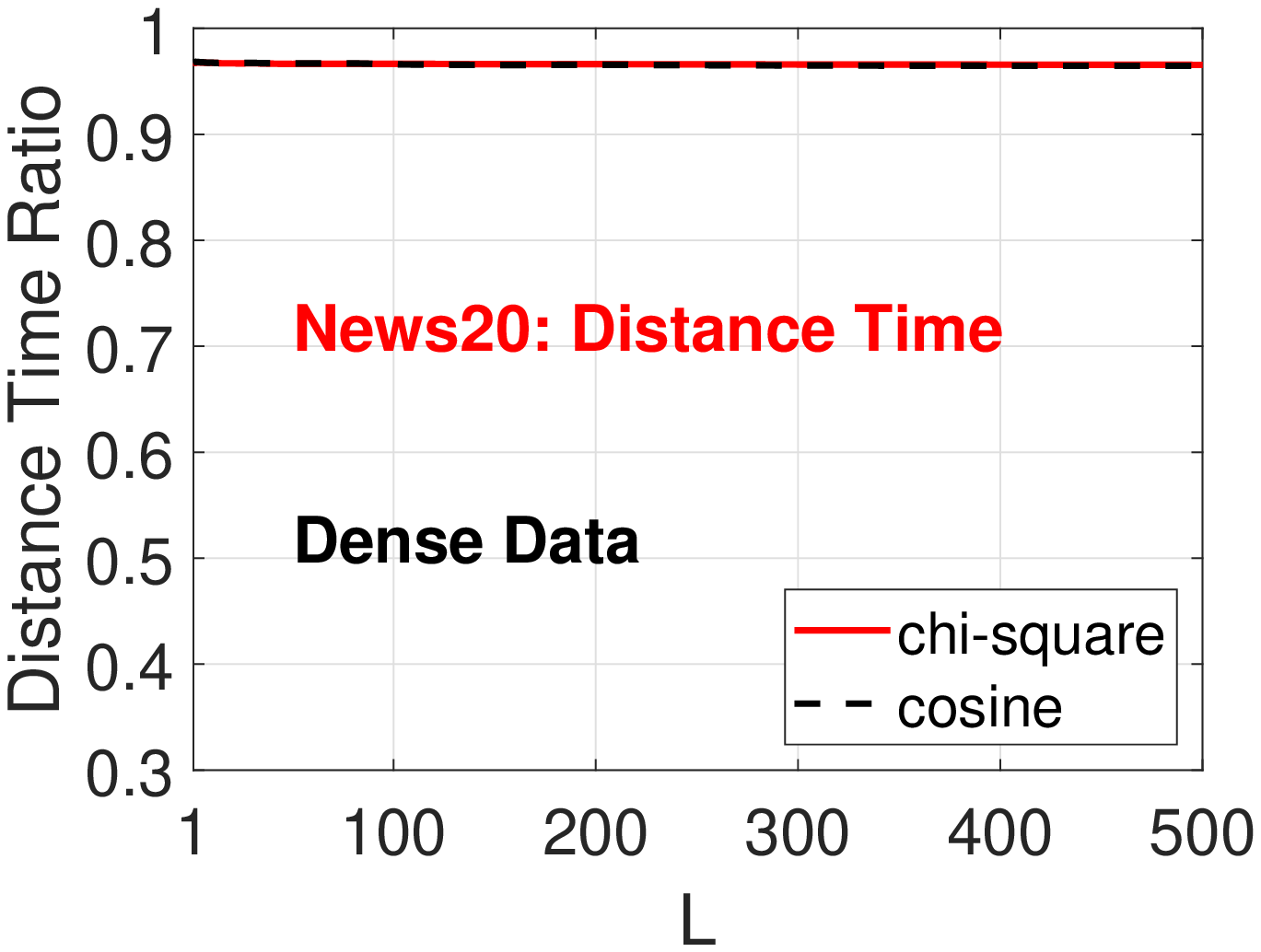}
}


    \caption{For Webspam, $>90\%$ of the  search time in HNSW is spent on computing similarities. For MNIST and News20, computing similarities  counts for $40\%\sim 50\%$ of the  search time. For News20, if we use the dense format, then computing similarities would take more than $95\%$ of the total time. }
    \label{fig:dist_time}
\end{figure}

\begin{figure}[b!]

\vspace{-0.1in}

\mbox{\hspace{-0.2in}
\includegraphics[width=1.85in]{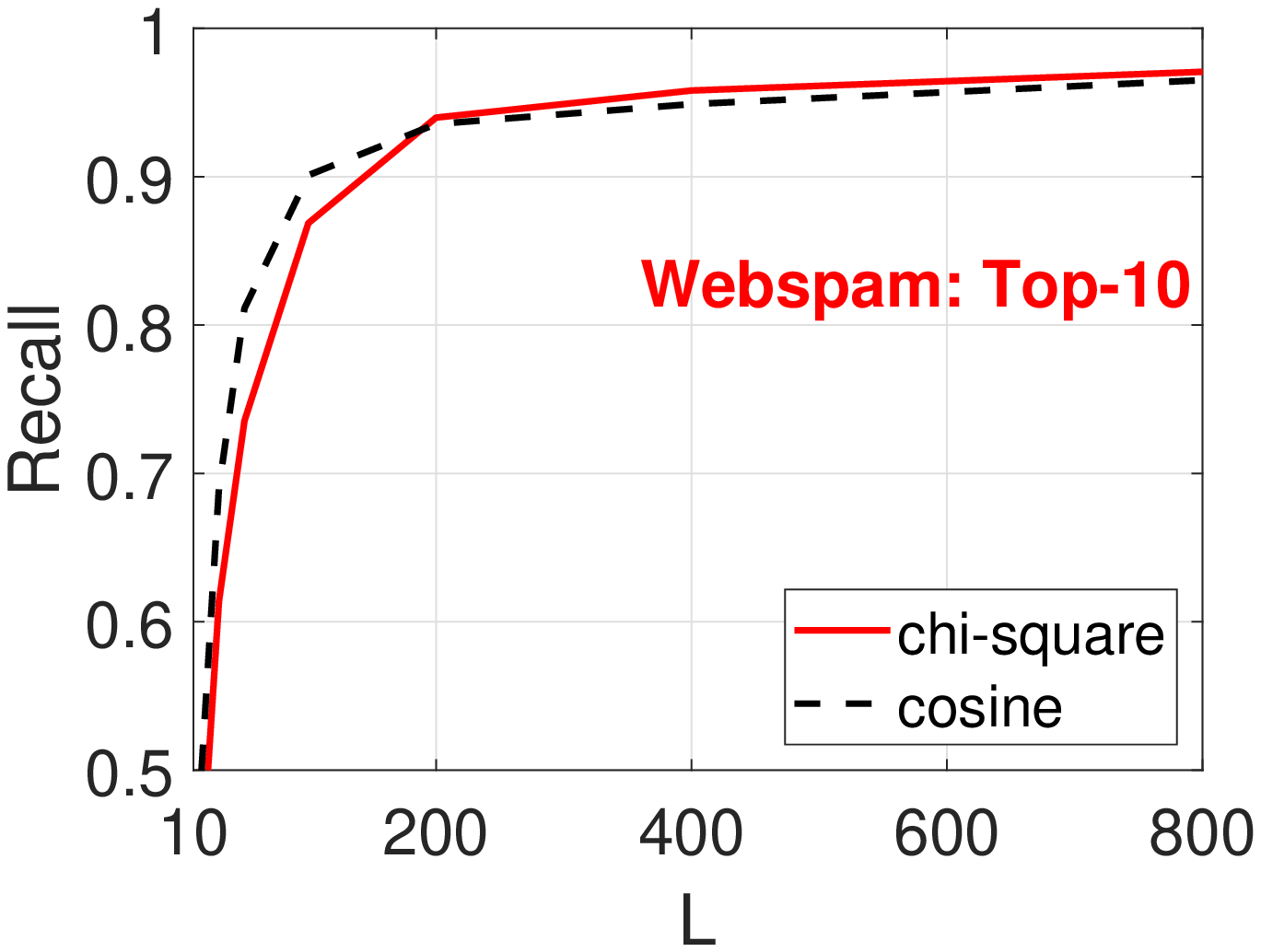}\hspace{-0.1in} \includegraphics[width=1.85in]{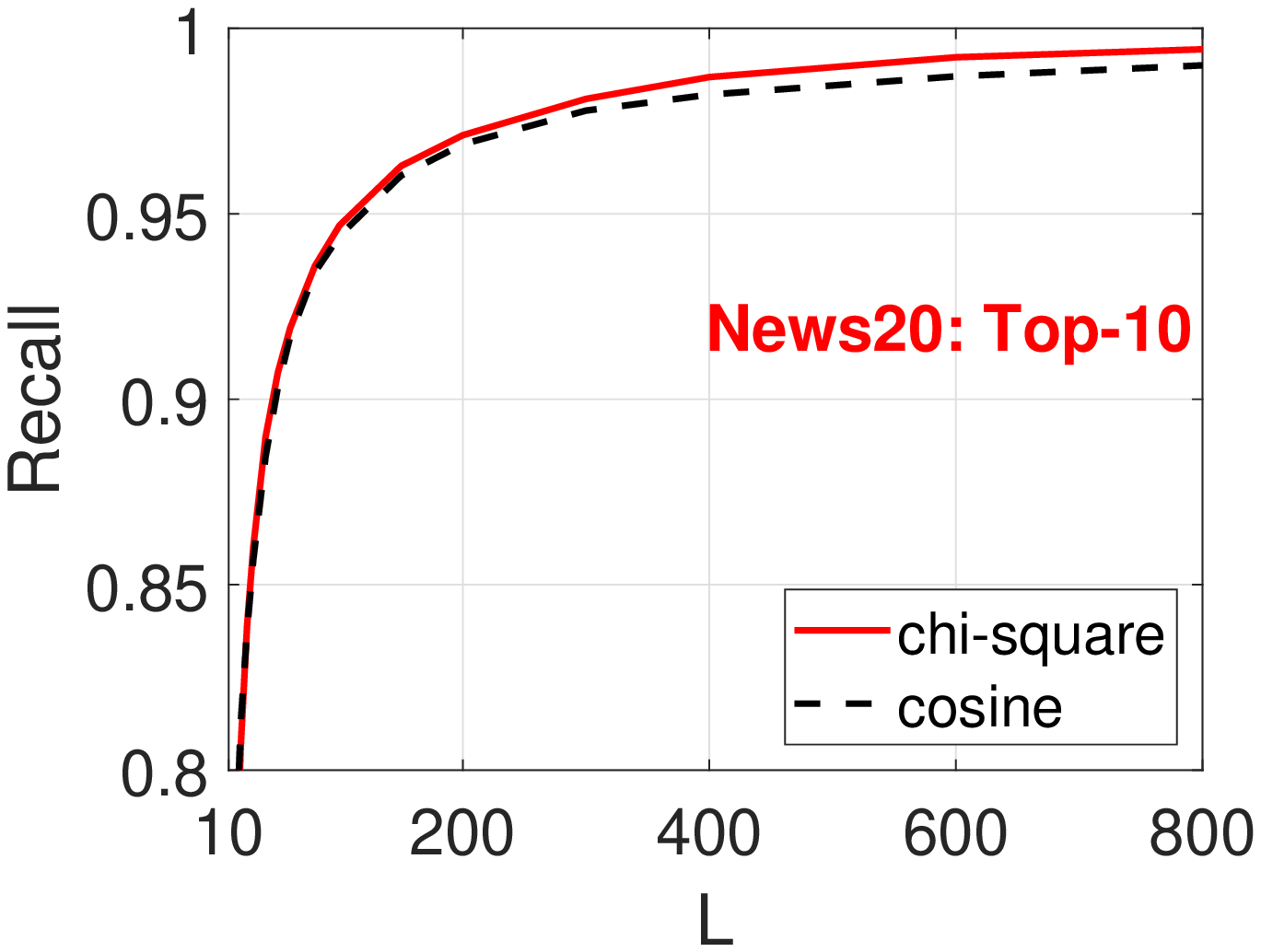}
}

\mbox{\hspace{-0.2in}
\includegraphics[width=1.85in]{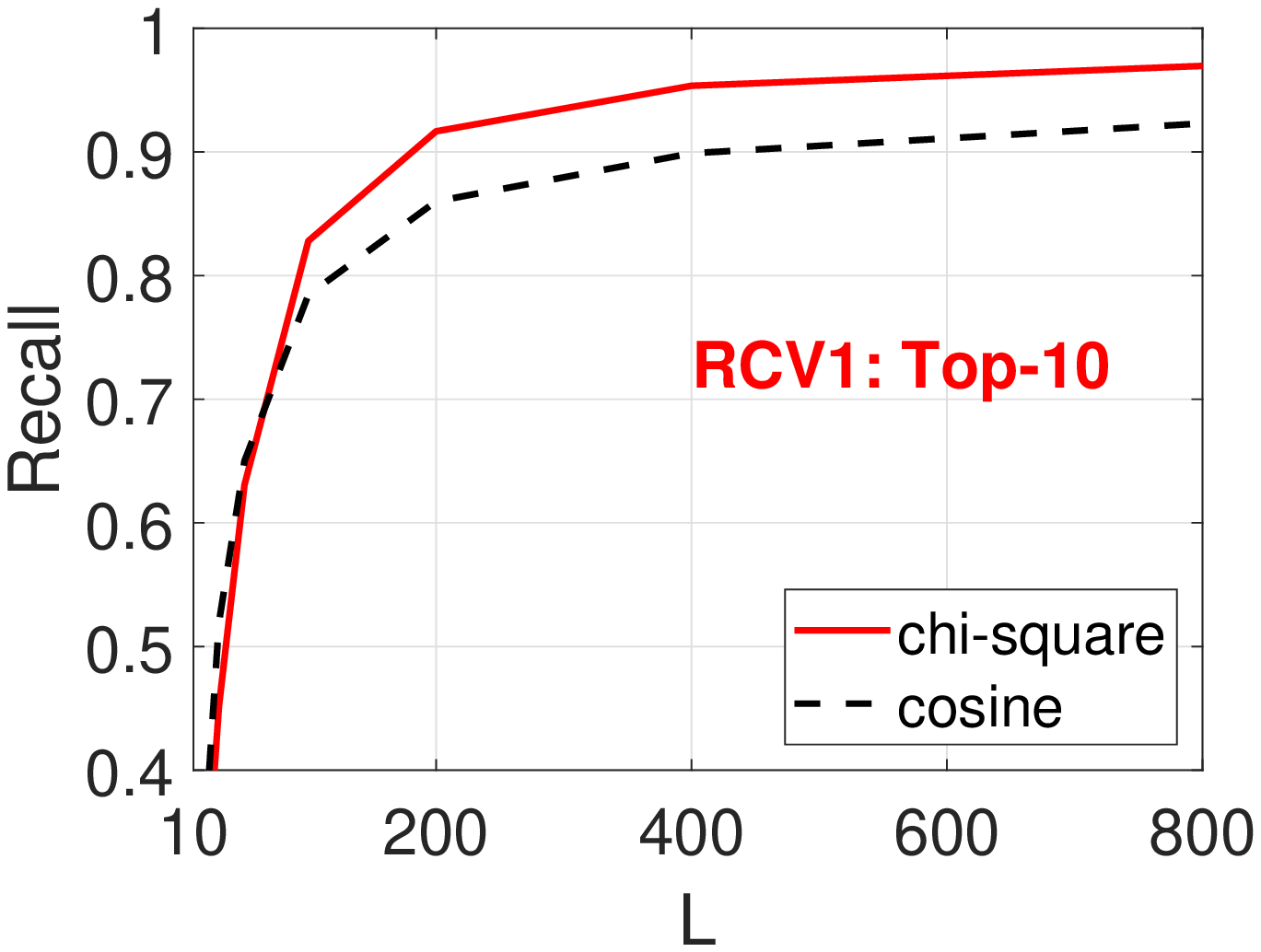}\hspace{-0.1in}\includegraphics[width=1.85in]{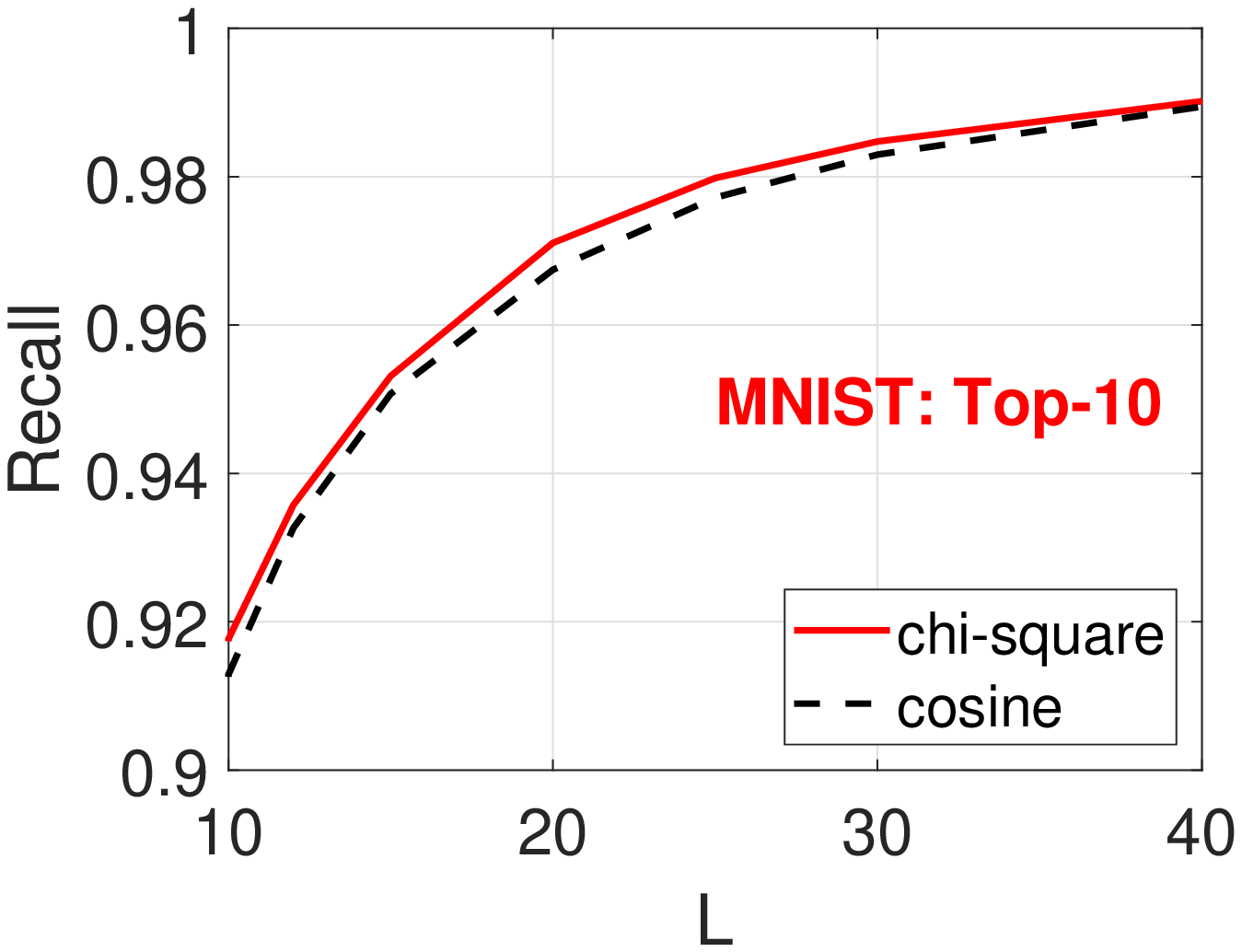}
}


    \caption{Recalls for retrieving top-10 similar vectors, for all four datasets, and for both  cosine and chi-square similarities, with respect to the HNSW parameter $L$.}
    \label{fig:recalls}
\end{figure}

Figure~\ref{fig:dist_time} reports the proportions of the times for computing similarities within the total search times. For Webspam, as the average number of nonzeros is high,  $>90\%$ of the search time is spent on computing similarities. For News20, the average number of nonzeros is  small and  only $40\%$ of the search time is spent on computing similarities. If we  use dense format for News20, then $>95\%$ of the search time would be spent on computing similarities.

\begin{figure}[t!]

\centering


\mbox{
    \includegraphics[width=1.85in]{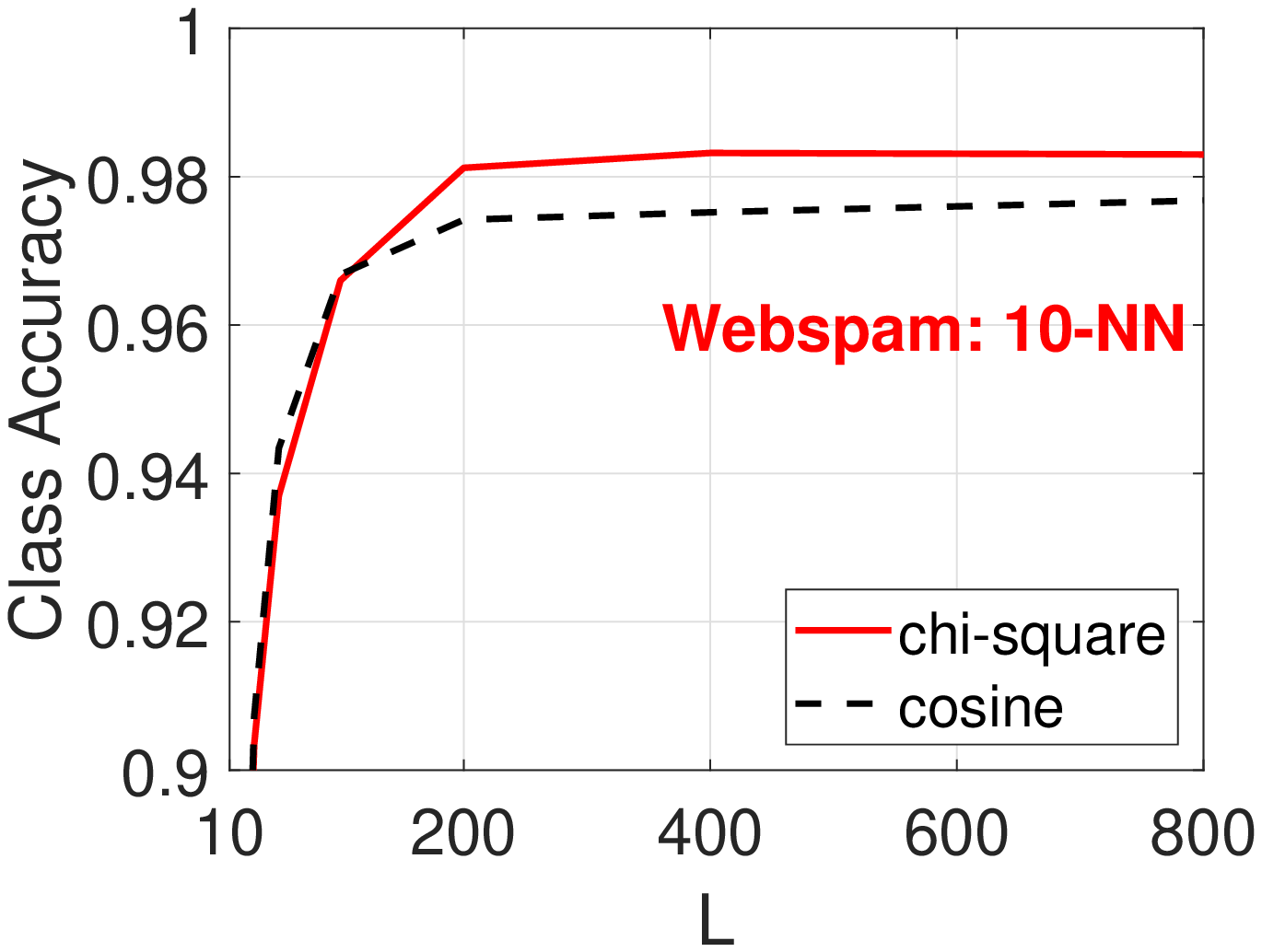}\hspace{-0.1in}
\includegraphics[width=1.85in]{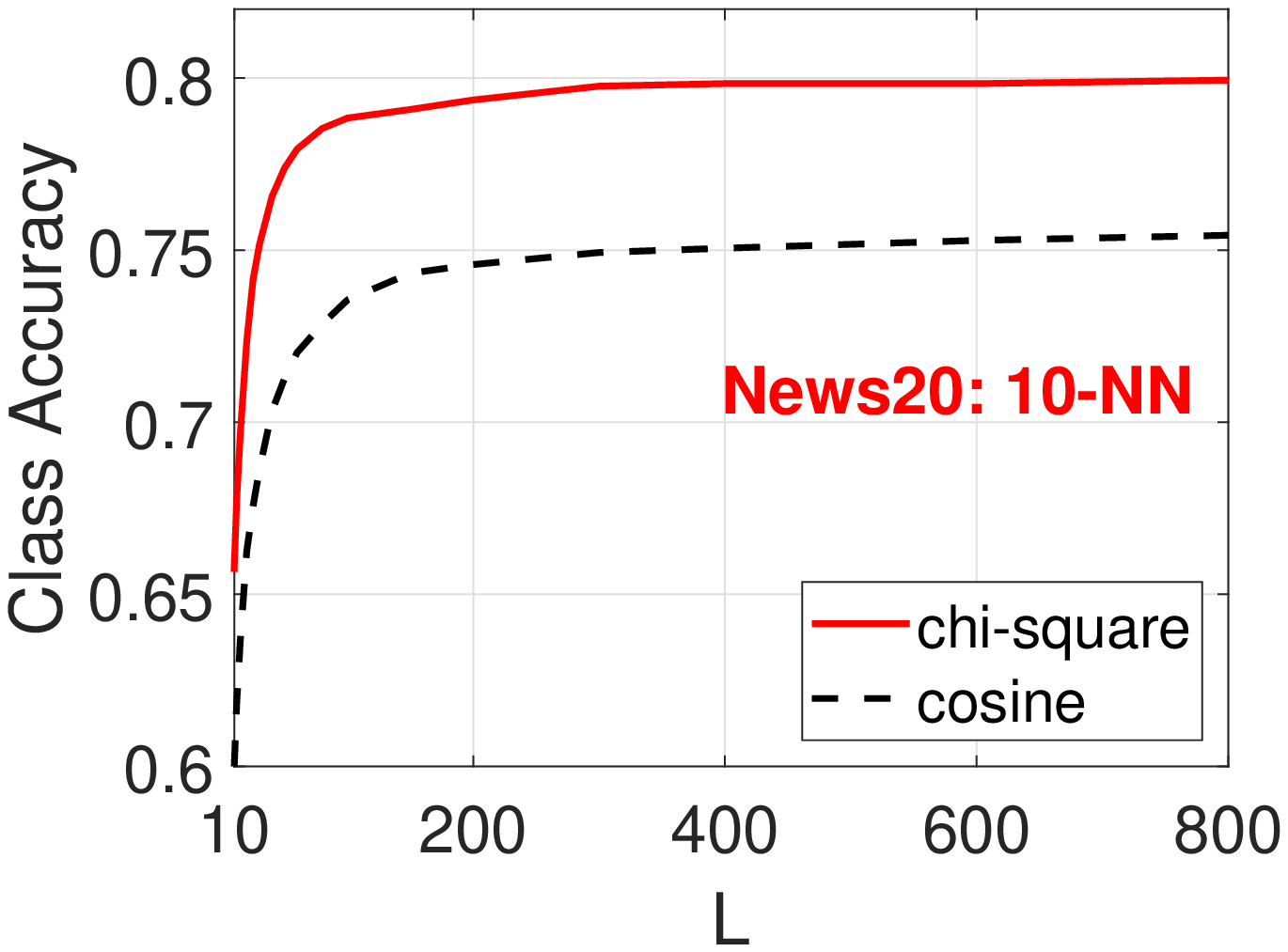}
}

\mbox{
    \includegraphics[width=1.85in]{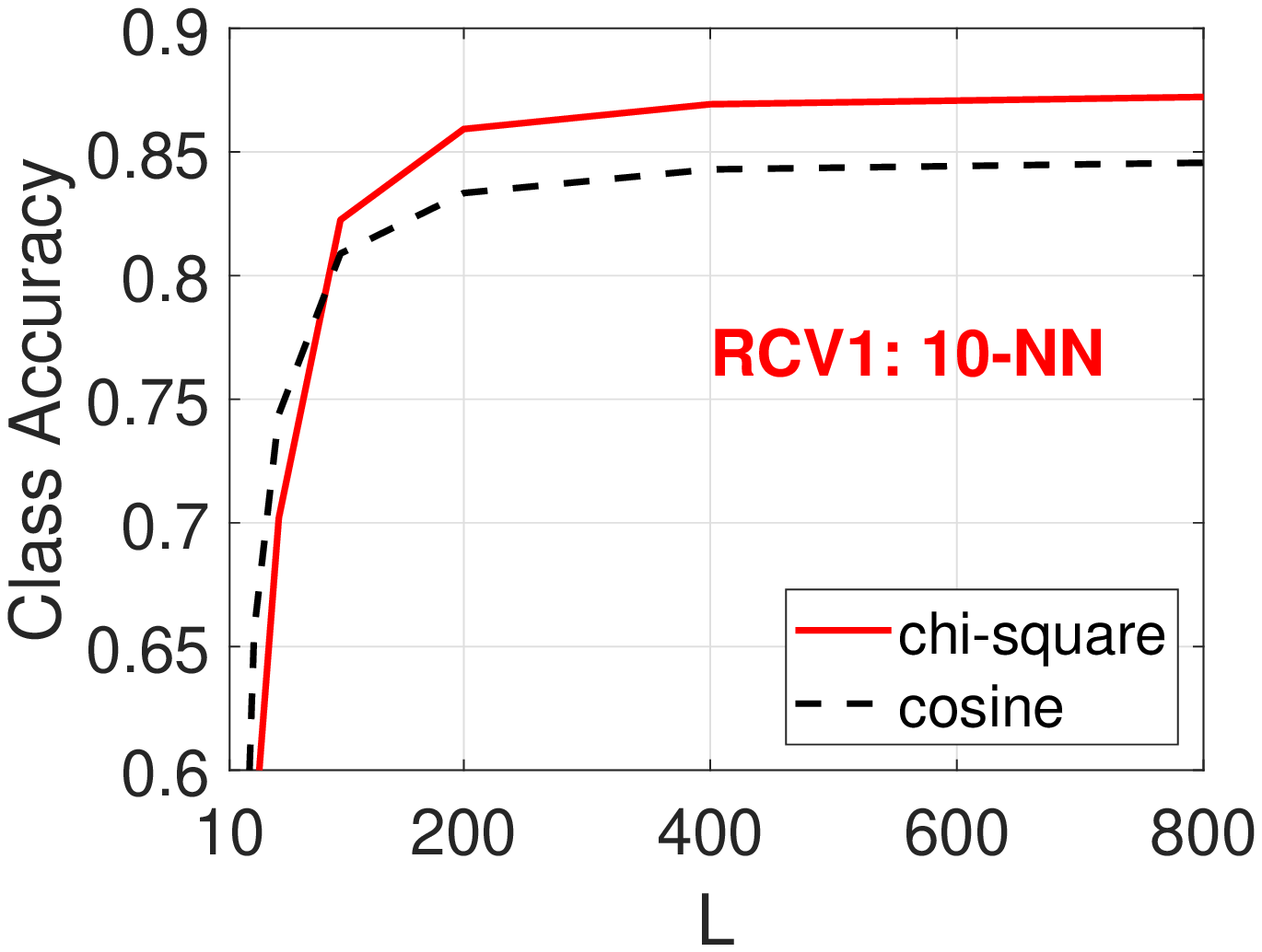}\hspace{-0.1in}
    \includegraphics[width=1.85in]{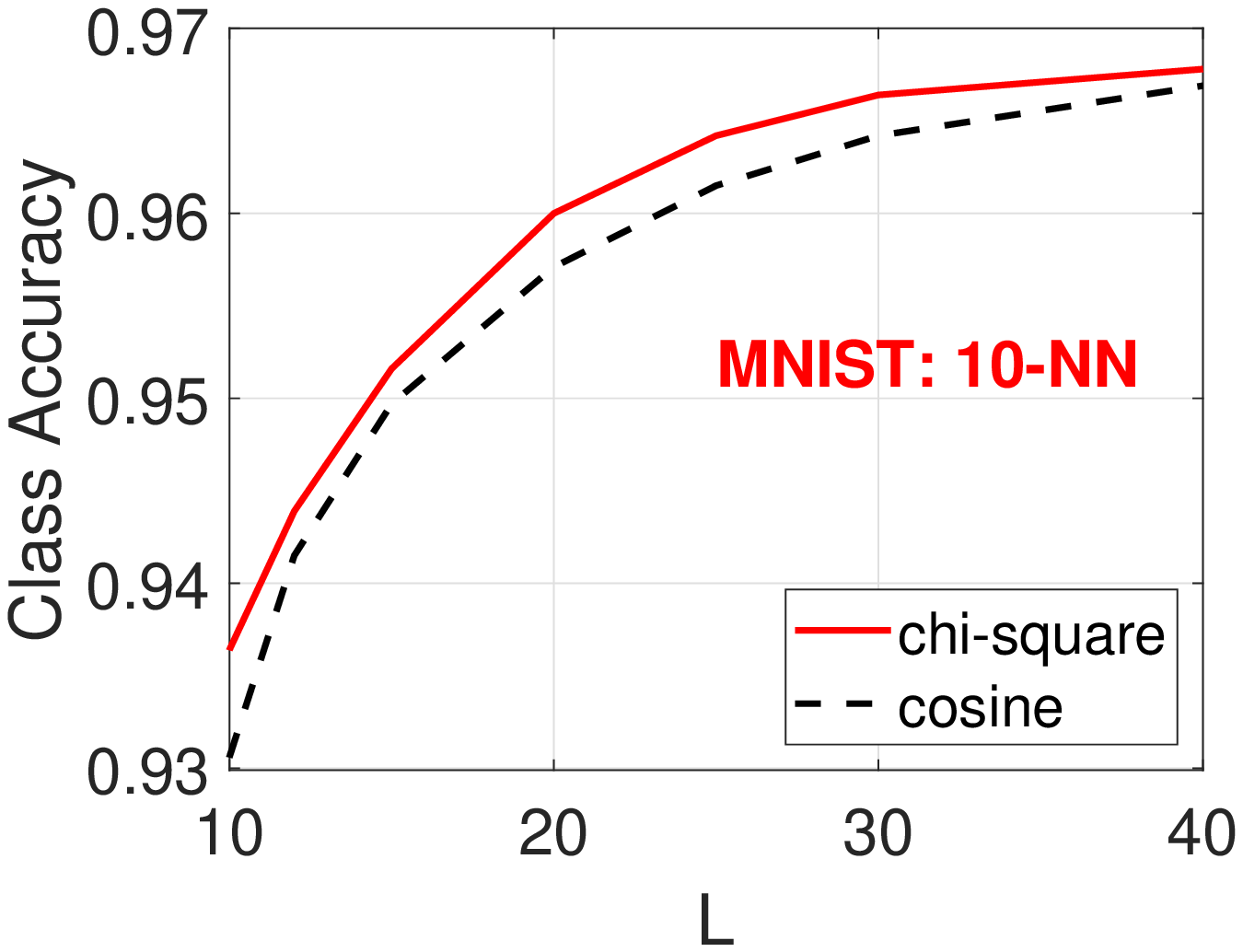}
}

\vspace{-0.1in}

    \caption{Classification accuracy  using the retrieved top-10  similar vectors (i.e., 10-NN), with respect to the HNSW parameter $L$. The chi-square performs better on these datasets.}
    \label{fig:class_acc}
\end{figure}

Figure~\ref{fig:recalls} presents the top-10 recalls for all four datasets with respect to $L$, for both the cosine and the chi-square similarities. Figure~\ref{fig:class_acc} reports the classification accuracy based using 10-NN (10 nearest neighbors) from the retrieved vectors. These results confirm that the chi-square similarity can be a good similarity measure to use for machine learning as well as retrieval.

\section{Sign Cauchy Random Projections}
\vspace{0.1in}

Before concluding the paper, we also report a solution for potentially further reducing the memory consumption of EBR in particular for the chi-square similarity. For example, for Webspam, even though the vectors are highly sparse, the absolute number of nonzeros is still high (3720). We propose to use ``sign cauchy random projections''~\citep{li2013sign} to reduce each vector to $k$ bits. Specifically, let $x_j = \sum_{i=1}^d u_i r_{ij}$ and $y_j = \sum_{i=1}^d v_i r_{ij}$, $j=1$ to $k$, where $\{r_{ij}\}$ is a random matrix with entries $r_{ij}$ sampled i.i.d. from the standard cauchy distribution. It was shown in~\citep{li2013sign} that the collision probability $P(sign(x_j) = sign(y_j))$ is proportional to the chi-square similarity. This is the foundation of ``sign cauchy random projections''.

\vspace{0.1in}

\begin{figure}[h!]

\mbox{\hspace{-0.2in}
    \includegraphics[width=1.85in]{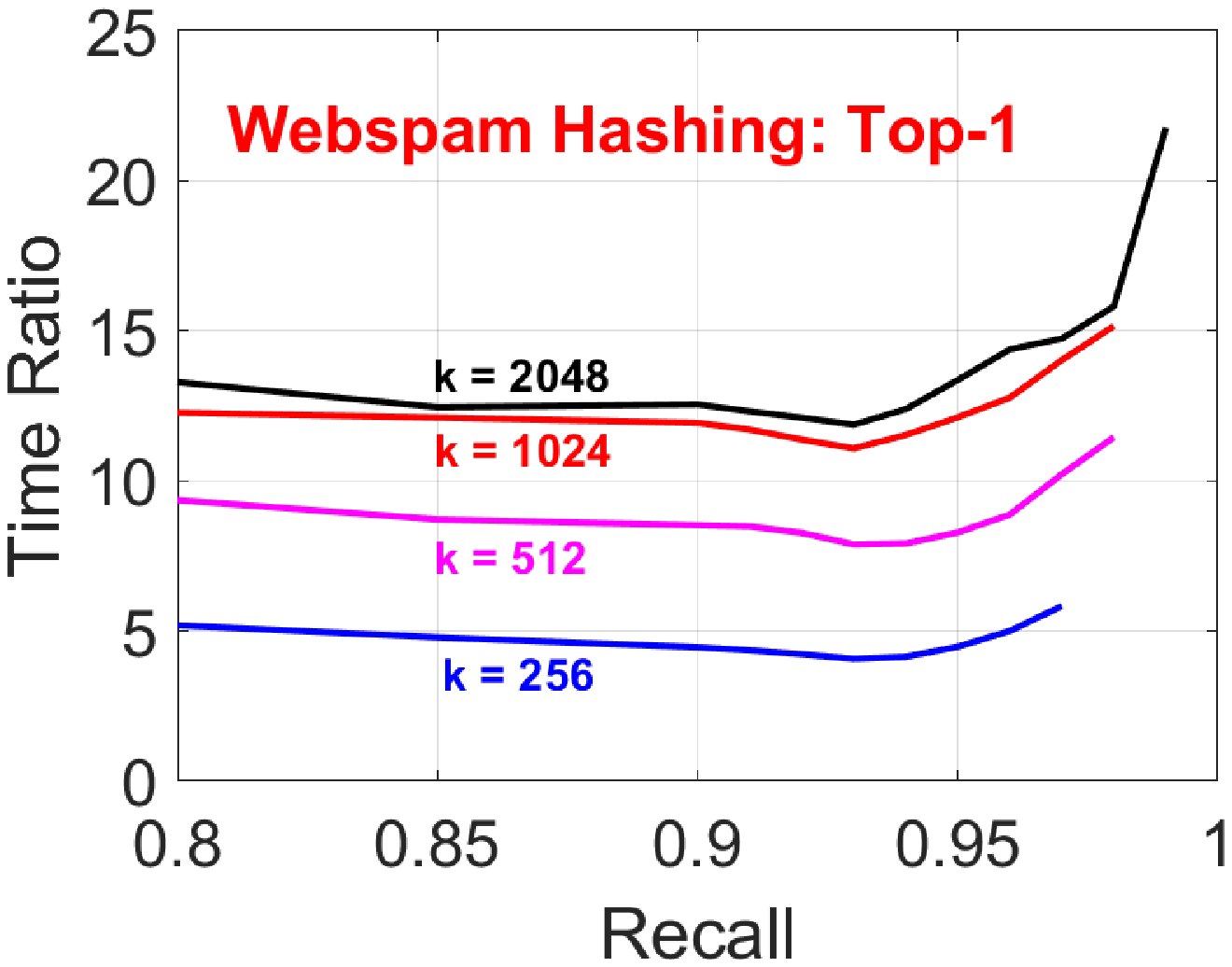}\hspace{-0.1in}
    \includegraphics[width=1.85in]{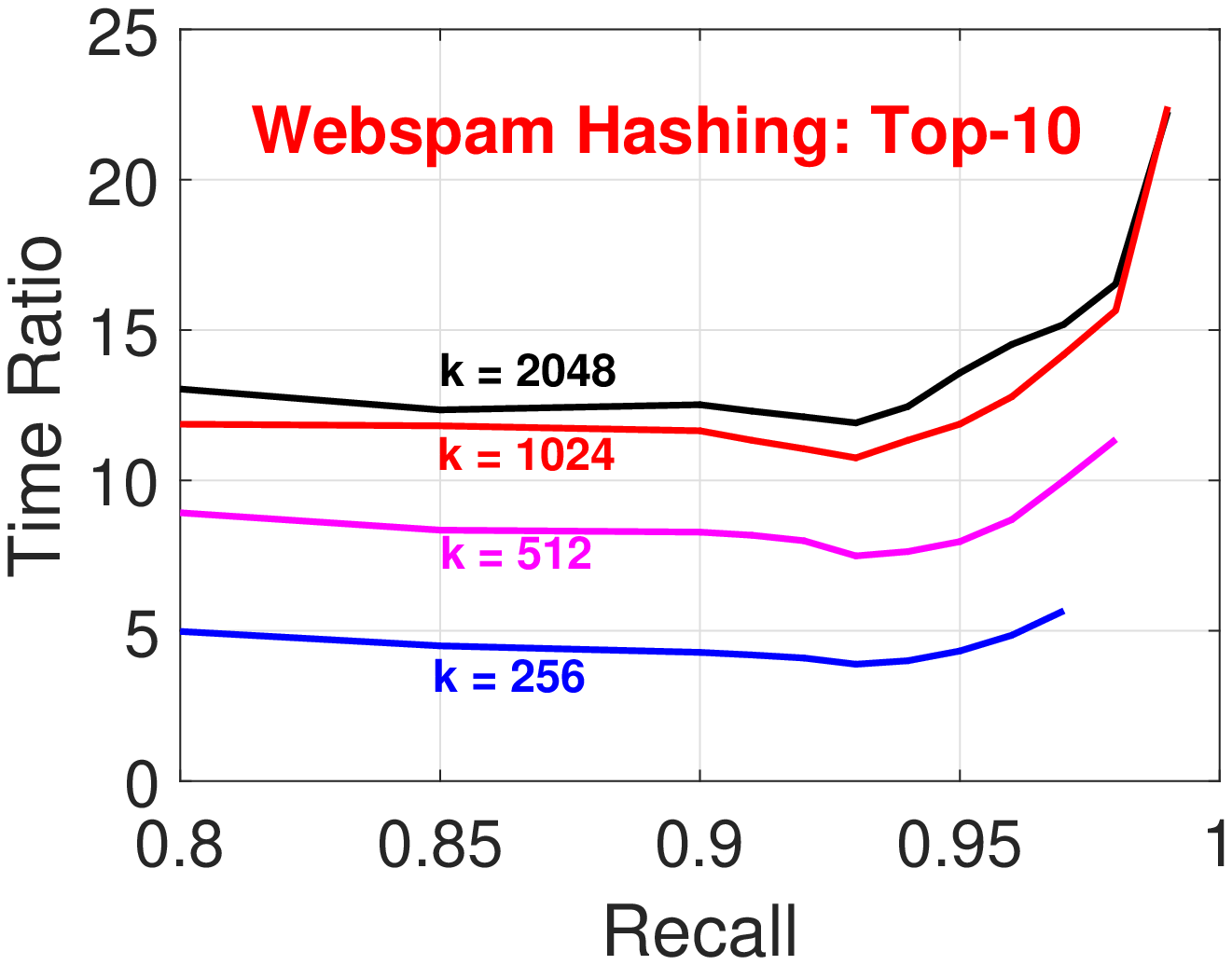}
}

\mbox{\hspace{-0.2in}
    \includegraphics[width=1.85in]{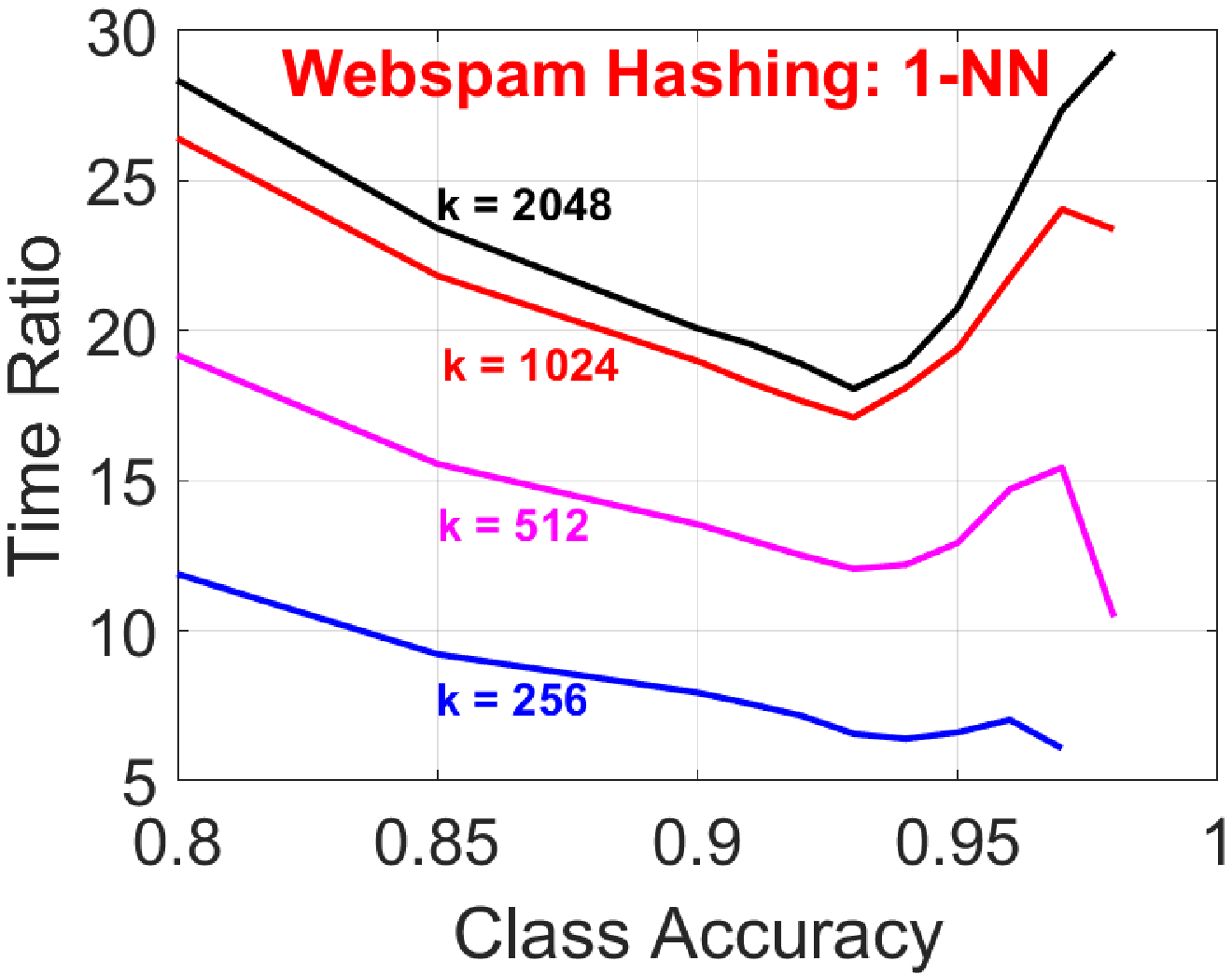}\hspace{-0.1in}
    \includegraphics[width=1.85in]{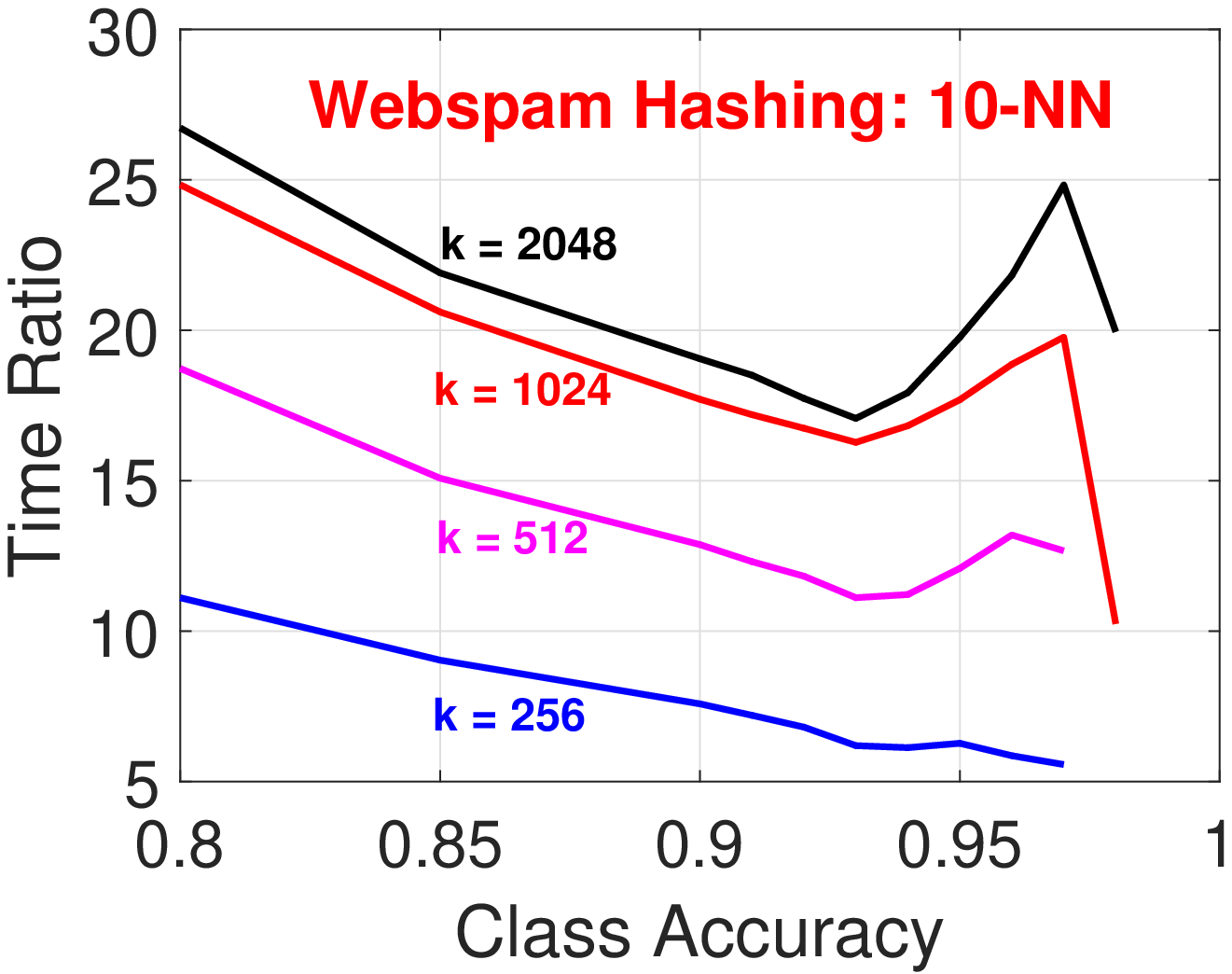}
}

\vspace{-0.1in}

    \caption{Ratios of similarity computing times: original chi-square over $k$ sign cauchy projections.  }
    \label{fig:cauchy}\vspace{-0.13in}
\end{figure}

Figure~\ref{fig:cauchy} reports the experiments on Webspam, which has 3720 nonzeros per vector. With $k=1024$ sign cauchy projections for estimating the original chi-square in HNSW,  we only need 1024/32 = 32 (4-byte) integers. To achieve the same 0.9 recall (respectively  0.9 classification accuracy), HNSW with the original data would need 12x (respectively 18x) more time for computing similarities. In addition, note that having greater $k$ yields better time for the same recall level. The reason is that a greater $k$ better approximates the original chi-square. Although it takes a slightly longer time to compute the hamming distance for greater $k$, a smaller $L$ is required to obtain the same recall, which still saves the entire searching cost.

\section{Conclusion}

The core contribution of this study is the integration of several techniques including the chi-square two-tower model, sign cauchy random projections (SignCRP), graph-based approximate near neighbor (ANN) search, and HNSW. These techniques are needed in part because the search data are often very sparse. We demonstrate these aspects through an industrial application in ads targeting.

\vspace{0.1in}

\noindent For many search applications, the data can be from the traditional ``handcrafted'' features, and data can also be generated from the trained deep learning models via (e.g.,) the two-tower models. In addition to the standard ``cosine two-tower'' model, in this study we also develop the ``chi-square two-tower'' model. Both models produce sparse embeddings when the ``ReLU'' activation function is used. The  chi-square similarity or chi-square loss function were very popular in the ``pre-deep-learning era'', for histogram-based features common in NLP and computer vision applications. It is a good exploration to bring the chi-square similarity to  deep learning.

\vspace{0.1in}

\noindent Approximate near neighbor (ANN) search is a critical component in modern recommender systems. Among many ANN algorithms, the graph-based methods such as HNSW or SONG (which is a GPU version of HNSW) become increasingly popular owing to the excellent performance. We focus on HNSW in this study. After the graph has been built, the major computational cost of HNSW is the evaluation of distance/similarity on the fly. Typically the  HNSW implementations assume dense data. Our study provides two solutions. The first (and obvious) solution is to use sparse representations for the storage as well as similarity computations.

\vspace{0.1in}

\noindent The second (and less obvious) solution is to apply ``sign cauchy random projections'' (SignCRP) to produce highly compact bits representations of the data and then leverage extremely efficient bit-wise operations to estimate the chi-square similarity (which is proportional to the number of matched bits). This strategy has been implemented and well-integrated into HNSW (and SONG). Experiments have confirmed the effectiveness of the proposed methods.


\balance
\bibliographystyle{ACM-Reference-Format}
\bibliography{refs_scholar}

\end{document}